\documentclass[letterpaper, 11pt]{article}

\usepackage[utf8]{inputenc}

\usepackage{amsmath}
\usepackage{amssymb}
\usepackage{amsthm}
\usepackage[font=small]{caption} \usepackage{epsfig}
\usepackage{geometry}
\usepackage{graphicx}
\usepackage{url}
\usepackage[colorlinks=true, citecolor=blue, filecolor=black, linkcolor=red, urlcolor=blue]{hyperref}
\usepackage{multirow}
\usepackage{makecell}
\usepackage{changes}
\usepackage{algorithm}
\usepackage{algpseudocode}
\usepackage[caption=false]{subfig}
\usepackage{soul} \setstcolor{red}
\setulcolor{blue}
\algdef{SE}[SUBALG]{Indent}{EndIndent}{}{\algorithmicend\ }%
\algtext*{Indent}
\algtext*{EndIndent}
\usepackage{rotating}
\usepackage{longtable}
\usepackage[mathlines]{lineno}
\usepackage[capitalise,nameinlink,noabbrev]{cleveref} 
\usepackage{cite}

\title{Bayesian Model Selection for Network Discrimination and Risk-informed Decision Making in Material Flow Analysis}

\author{Jiankan Liao\footnote{\href{mailto:jkliao@umich.edu}{jkliao@umich.edu}, Ph.D. student, Mechanical Engineering, University of Michigan, Ann Arbor, MI 48109.},
Sidi Deng\footnote{\href{mailto:sidideng@umich.edu}{sidideng@umich.edu}, Postdoctoral researcher, Mechanical Engineering, University of Michigan, Ann Arbor, MI 48109.},
Xun Huan\footnote{\href{mailto:xhuan@umich.edu}{xhuan@umich.edu}, Associate Professor, Mechanical Engineering, University of Michigan, Ann Arbor, MI 48109. \href{https://uq.engin.umich.edu}{https://uq.engin.umich.edu}}, and
Daniel Cooper\footnote{Corresponding author: \href{mailto:drcooper@umich.edu}{drcooper@umich.edu}, Associate Professor, Mechanical Engineering, University of Michigan, Ann Arbor, MI 48109.}
}

\begin{document}


\maketitle

\begin{abstract} 

Material flow analyses (MFAs) provide insight into supply chain level opportunities for resource efficiency. MFAs can be represented as networks with nodes that represent materials, processes, sectors or locations. MFA network structural uncertainty (i.e., the existence or absence of flows between nodes) is pervasive and can undermine the reliability of the flow predictions. This article investigates MFA network structure uncertainty by proposing candidate node-and-flow structures and using Bayesian model selection to identify the most suitable structures and Bayesian model averaging to quantify the parametric mass flow uncertainty. The results of this holistic approach to MFA uncertainty are used in conjunction with the input-output (I/O) method to make risk-informed resource efficiency recommendations. These techniques are demonstrated using a case study on the U.S. steel sector where 16 candidate structures are considered. Model selection highlights 2 networks as most probable based on data collected from the United States Geological Survey and the World Steel Association. Using the I/O method, we then show that the construction sector accounts for the greatest mean share of domestic U.S. steel industry emissions while the automotive and steel products sectors have the highest mean emissions per unit of steel used in the end-use sectors. The uncertainty in the results is used to analyze which end-use sector should be the focus of demand reduction efforts under different appetites for risk. This article’s methods generate holistic and transparent MFA uncertainty that account for structural uncertainty, enabling decisions whose outcomes are more robust to the uncertainty.

\end{abstract}


\section{Introduction}

Material flow analyses (MFAs) are typically represented as directed graphs to track the flow of a resource (e.g., aluminum) along a supply chain \cite{Brunner16}. As described by Cullen and Cooper \cite{Cullen22}, MFAs are essential for evaluating the potential environmental impacts of material (resource) efficiency because opportunities for such efficiency are dispersed across the supply chain and product life cycle. MFAs elucidate the connections between material production, yields in manufacturing processes, and end-user demand for products. To accurately gauge the impact on emissions, encompassing both upstream and downstream effects, MFAs are necessary since emission-intensive processes (such as clinker production in cement kilns) may be spatially and temporally distant from the intervention itself (e.g., improved bridge maintenance for prolonged concrete lifespan). Through MFA, mitigation strategies from various points within the material system can be compared on an equal footing.

The data used to construct an MFA are often sparse, noisy, and diverse; e.g., data pertaining to the mass flow between two processes might be expressed as a percentage of the sum of all the mass flows into a group of processes or alternatively in relation to flows elsewhere in the network \cite{Kopec16} (see section S1 of the Supporting Information (SI) for more details on the different forms of MFA data). Given the poor data quality typically used to construct an MFA, making environmentally motivated decisions and policies without uncertainty and confidence measures may lead to reduced or even negative environmental benefits. For these reasons, it has in recent years been increasingly recognized that it is important to quantify MFA uncertainty to enable informed decision- and policy-making~\cite{Schwab18,Graedel19,Cullen22}.

For any MFA problem, the mass flow uncertainty is the combined effect of the parametric uncertainty (e.g., the uncertain allocation of flows through one node that are destined to another node) given a certain network structure (i.e., the nodes and connections between the nodes that define the structure of the directed graph) and the network structure uncertainty itself; i.e., the presence or absence of nodes and/or connections between nodes. Typical origins of parametric uncertainty include a lack of data on mass flows, mass flow measurement/data record error, and imputation; e.g., calculating the iron ore produced in Minnesota based on nationwide statistics and estimates of Minnesota's market share. Network structure uncertainty may originate from the MFA practitioner's lack of expertise regarding the material system in question or conflicting evidence on the correct sequence of processes in the supply chain. Network structure uncertainty may be exacerbated if the true MFA network structure has changed in recent years and/or varies across regions.

A popular tool for quantifying parametric uncertainty in MFA is the STAN open-source package \cite{Cencic08}. By representing collected data inputs using probability density functions (PDFs), STAN reconciles the collected data through least square fitting and uses error propagation to determine the mass flow uncertainty \cite{Cencic16}. However, such PDFs are often not available; e.g., no error bars or PDFs are published alongside the U.N. Comtrade Database's mass flow statistics \cite{UNcomtrade12}.
Furthermore, the STAN package does not allow ``multiple data records to be directly considered for an individual flow variable''~\cite{Zhu19}. Alternatively, Bayesian inference, a general probabilistic approach to uncertainty quantification, has been effectively used to handle MFA parametric uncertainty \cite{Gottschalk10,Lupton18,Dong23,Wang22}. 
Following Bayes' rule, an initial probability distribution (the ``prior'') representing the starting uncertainty in MFA variables is updated (conditioned) based on observed material flow data, resulting in a new distribution (the ``posterior'') that reflects the refined uncertainty informed by the data (see, e.g., \cite{Berger85,Jaynes03,Sivia06, Bertsekas08,VonToussaint11}). 
The Bayesian framework is powerful because it allows the incorporation of domain knowledge through the prior, integrates data from multiple sources, and derives a justified level of uncertainty even when data is sparse or noisy. Moreover, Bayes' rule can be applied iteratively, allowing a sequential  updating of the MFA uncertainty as new data is collected, with the posterior from one iteration acting as the prior for the next.

Network structure uncertainty is typically not acknowledged in MFAs and has not been rigorously studied. While observation of a large discrepancy between collected and reconciled MFA data may suggest a missing flow in the network \cite{Anspach24}, there is currently no systematic method that quantifies the network structure uncertainty in MFA.
However, network structure uncertainty is considered a form of model uncertainty, and model discrimination under a Bayesian framework has been explored elsewhere, including in related problems such as determining whether MFA parameters change over multiple years \cite{Dong23} and in predicting future material demand \cite{Bhuwalka23}. Below, we first review the existing work on MFA network structural uncertainty (\cref{ss:lr_model}), the existing application of Bayesian parameter inference in MFA (\cref{ss:lr_bayesian}), and then the scope of this paper to extend the Bayesian approach in MFA to account for network structure uncertainty (\cref{ss:scope}).

\subsection{Previous work on network structure uncertainty in MFA}
\label{ss:lr_model}

Existing studies for uncertainty quantification in MFA are largely based on the assumption that the underlying network structure is correct. However, even if not acknowledged, there is often network structure uncertainty in MFAs. This is because MFAs often model complex supply chains for which detailed knowledge on each aspect of the MFA is dispersed across stakeholders and likely unavailable to the MFA practitioner. Material flow data recorded in the literature and by statistical agencies may also be simplified, mislabelled, or misinterpreted in a manner that suggests flows exist between nodes where there are none in reality and vice versa. Supply chain structures for nominally identically materials may also vary by region, introducing further uncertainty. For example, Klinglmair \textit{et al.} \cite{Klinglmair16} show that the structure of the phosphorus material flow in Denmark is very different from that in Austria.

To the authors' knowledge, no published research so far has focused on MFA network structure uncertainty. Research pertaining to the evaluation of network structure more generally includes Chatterjee \textit{et al.}'s \cite{Chatterjee23} study on supply chain resilience using graph-theoretic metrics and the concept of ``window of vitality'' from the ecology field aimed at balancing redundancy and efficiency. Elsewhere, Schwab and Rechberger \cite{Schwab18} build on the network resilience theory developed by Ulanowicz \textit{et al.} \cite{Ulanowicz09} and the notion of ``information defect'' defined by Schwab \textit{et al.} \cite{Schwab16}. They define an MFA system complexity metric as a function of the number of nodes and connections, as well as an MFA data size and quality metric. By comparing these two metrics, Schwab and Rechberger derive a system property that provides an ordinal measure of the extent to which a system is known (0--100\%). For example, a simple MFA network with few nodes and connections and noiseless data collection on each mass flow would result in the MFA ``system [being] known to an extent of [100\%]'' \cite{Schwab18}. The system property defined by Schwab and Rechberger allows a quantitative measure of the state of knowledge on different MFAs or the improving state of knowledge during the incremental development of a given MFA. However, neither Chatterjee \textit{et al.} nor Schwab and Rechberger provide a measure of a given network structure being the correct structure for a given system.

\subsection{Previous work on Bayesian inference in MFA}
\label{ss:lr_bayesian}

Bayesian parameter inference has previously been used to quantify mass flow uncertainty but assuming the network structure, defined by the MFA practitioner, is correct. Gottschalk \textit{et al.} \cite{Gottschalk10} were the first to use Bayesian inference for this purpose, quantifying the uncertainty of nano-TiO\textsubscript{2} particle releases into the environment in Switzerland. They introduced the concept of transfer coefficients, also known as allocation fractions (i.e., the fraction of the total flow through a node that is destined to flow to a given downstream node), to model an MFA network as a linear system using matrix algebra, automatically guaranteeing mass balance as long as all transfer coefficients emanating from a node sum to 1. Instead of forming priors on the mass flows directly, they applied uni-variate prior distributions on individual transfer coefficients using historical data.
Later, Lupton and Allwood \cite{Lupton18} adopted a multi-variate Dirichlet distribution to model the prior distribution jointly on all transfer coefficients emanating from a given node. The use of Dirichlet distribution provides a flexible way to construct the prior distribution on a node's transfer coefficients that automatically sums to unity. Lupton and Allwood demonstrated their method by remapping the 2008 global steel flow from Cullen \textit{et al.} \cite{Cullen12} using mainly uniform Dirichlet priors and assuming a constant noise level on all collected MFA data of $\pm 10\%$. More recently, Dong \textit{et al.} \cite{Dong23} studied expert elicitation \cite{OHagan05} and data noise learning for MFA using Bayesian inference. They demonstrated how informed multivariate MFA priors can be derived using expert elicitation and how the MFA data noise level can be inferred concurrently with the MFA parameters by modeling the noise as a random variable. Elsewhere, Wang \textit{et al.} \cite{Wang22} took a different approach where instead of parameterizing the MFA 
using transfer coefficients, they assigned priors on mass flows and process stocks directly, and imposed mass balance constraints 
through a violation penalty in the likelihood function.
They demonstrated their method on two case studies, an aluminum system in 2009 based on Liu \textit{et al.} \cite{Liu13} and a zinc system from 1994 to 1998 based on Graedel \textit{et al.} \cite{Graedel05}. 

Several studies have compared the performance of different Bayesian setups in MFA related problems; e.g., determining the consistency of process yields across multiple years \cite{Dong23} and estimating income and price elasticity of future material demand \cite{Bhuwalka23}. While Dong \textit{et al.} \cite{Dong23} did not address network structure uncertainty, they did utilize Bayes factor (a quantitative metric for model selection that, as we will show, is useful for capturing network structure uncertainty) to justify using MFA data from multiple years to enhance learning of MFA data noise parameters. Bhuwalka \textit{et al.} \cite{Bhuwalka23} used Bayesian hierarchical regression to model copper demand in five regions and sectors as a function of price and income, and compared the model fitting results with those generated from an un-pooled model (individual models for individual regions and sectors) and a fully pooled model (one global model for any region and sector). The result from the hierarchical model showed better uncertainty-reduction than the un-pooled model, while capturing copper demand characteristics region- and sector-wise compared with the fully-pooled model. Despite the advantages of applying a Bayesian framework to MFA, Bornhoft \textit{et al.} \cite{Bornhoft21} pointed out the increased modeling effort as a potential drawback.

\subsection{Scope of this paper}
\label{ss:scope}

The key contributions of this paper are 1) applying Bayesian model selection to quantify the network structure uncertainty and calculate the probability of the candidate structures using collected MFA data, 2) deriving individual mass flow PDFs that account for both MFA data noise and network structure uncertainty, and 3) utilizing the mass flow uncertainty results for informed decision making on demand reduction for decarbonization. \Cref{s:formulation} introduces a stylized MFA model example to guide the reader through the methodology. \Cref{s:USGS} demonstrates the method using a case study on U.S. annual steel flow. Finally, \cref{s:discussion} concludes with lessons learned from the case study and potential future work.
\section{Methodology}
\label{s:formulation}

\begin{figure}
    \centering
    \includegraphics[width=0.69\textwidth,height=\textheight,keepaspectratio]{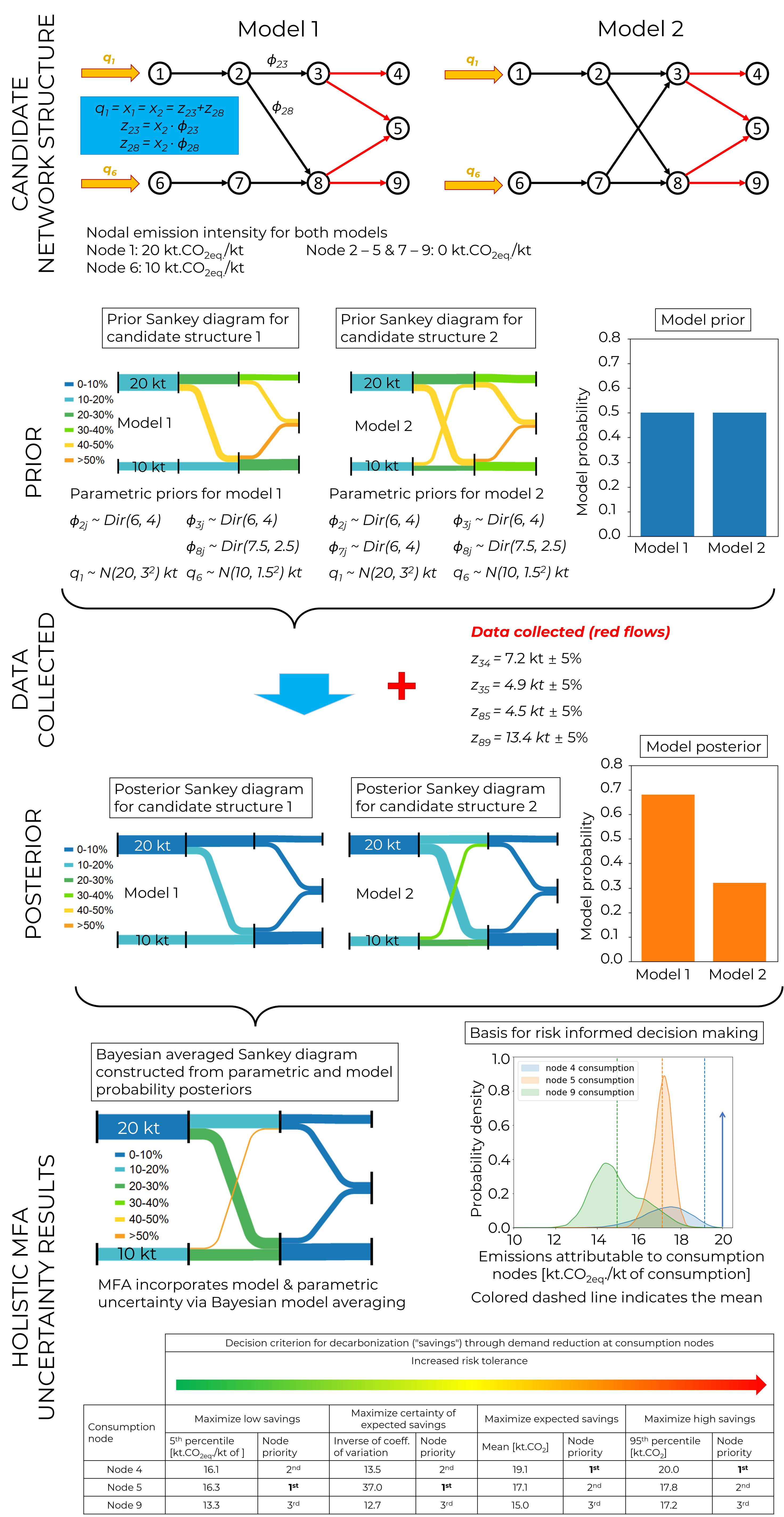}
    \caption{Demonstration of the MFA model selection and risk-informed decarbonization decision making through demand reduction procedure using a simple MFA model example. $Dir(\cdot)$ denotes the Dirichlet distributions with the corresponding hyper-parameters.
    }
    \label{f:toy_model}
\end{figure}

We first introduce the linear system to formulate the MFA network (\cref{ss:MFA_mathematical}). The Bayesian framework is then established to quantify the MFA parametric (\cref{ss:parametric_uncertainty}) and network structure uncertainty (\cref{ss:network_structure_uncertainty}). The final mass flow uncertainty then encompasses
both the parametric uncertainty under a given network structure, and the network structure uncertainty (\cref{ss:mass_flow_uncertainty}). Finally, we introduce common decision-making metrics and how to use the quantified mass flow uncertainty to guide decision making for resource efficiency (\cref{ss:decision_making}). \Cref{f:toy_model} gives an example of the overall procedure and is referred to throughout this section.

\subsection{A mathematical representation of MFA}
\label{ss:MFA_mathematical}

As described in previous work (e.g., \cite{Dong23}), an MFA can be illustrated using a directed graph, as depicted in \cref{f:toy_model} (top). The nodes in the graph, numbered 1, 2, \ldots, $n_p$, represent a total of $n_p$ processes, products, or locations. Each directed edge between two nodes signifies the mass flow of material from one process to another.

At the heart of MFA is the principle of mass conservation, which mandates that the total mass of material flowing into each node (total input) must equal the total mass of material flowing out of each node (total output). We represent the total input (or equivalently, total output) flow for node $i$ as $x_i$. The flow along an edge from node $i$ to node $j$ is then given by $z_{ij}= \phi_{ij} x_i$, where $\phi_{ij} \in [0,1]$ represents the allocation fraction of node $i$'s total outflow directed towards node $j$ ($\phi_{ij}$ = 0 if there is no flow from node $i$ to node $j$). Consequently,
\begin{align}
     \sum_{i=1}^{n_p}\phi_{ij}x_{i}=x_{j} \qquad \text{and} \qquad
    \sum_{j=1}^{n_p}\phi_{ij} = 1. 
\end{align}
We suggest using the allocation fractions ($\phi_{ij}$) as model parameters instead of the direct mass flow values. This is because, as explained by Gottschalk \textit{et al.} \cite{Gottschalk10}, the allocation fractions provide a convenient method of expressing and enforcing the mass balance relationships for the entire MFA in the form of a linear system. For example, the mass balance equations for the MFA model 1 ($M_1$) illustrated in \cref{f:toy_model} (top) can be formulated as:
\begin{equation}
    \label{e:matrix}
    \underbrace{
        \begin{bmatrix}
            1 & 0 & 0 & 0 & 0 & 0 & 0 & 0 & 0 \\
            -\phi_{12} & 1 & 0 & 0 & 0 & 0 & 0 & 0 & 0 \\
            0 & -\phi_{23} & 1 & 0 & 0 & 0 & 0 & 0 & 0 \\
            0 & 0 & -\phi_{34} & 1 & 0 & 0 & 0 & 0 & 0 \\
            0 & 0 & -\phi_{35} & 0 & 1 & 0 & 0 & -\phi_{85} & 0 \\
            0 & 0 & 0 & 0 & 0 & 1 & 0 & 0 & 0 \\
            0 & 0 & 0 & 0 & 0 & -\phi_{67} & 1 & 0 & 0 \\
            0 & -\phi_{28} & 0 & 0 & 0 & 0 & -\phi_{78} & 1 & 0 \\
            0 & 0 & 0 & 0 & 1 & 0 & 0 & -\phi_{89} & 1
        \end{bmatrix}
    }_{\mathbb{I}-\Phi^{\top}}
    \underbrace{
        \begin{bmatrix}
            x_1 \\
            x_2 \\
            x_3 \\
            x_4 \\
            x_5 \\
            x_6 \\
            x_7 \\
            x_8 \\
            x_9
        \end{bmatrix}
    }_{x}
    =
    \underbrace{
        \begin{bmatrix}
            q_1 \\
            0 \\
            0 \\
            0 \\
            0 \\
            q_6 \\
            0 \\
            0 \\
            0
        \end{bmatrix}
    }_{q},
\end{equation}
where $\mathbb{I}$ denotes the $n_p \times n_p$ identity matrix, $\Phi \in \mathbb{R}^{n_p \times n_p}$ is the adjacency matrix with entries being the allocation fractions $\phi_{ij}$, $x \in \mathbb{R}^{n_p}$ is the vector collecting all nodal mass flows, and $q \in \mathbb{R}^{n_p}$ is the vector catching any external inflows $q_i$ to the network (see \cref{f:toy_model}, top; e.g., aluminum imports in a country-level aluminum MFA).

Given $\Phi$ and $q$, the model prediction for all nodal mass flows can be solved as:
\begin{equation}
\label{e:node}
    x=(\mathbb{I}-\Phi^{\top})^{-1}q.
\end{equation}
The term $(\mathbb{I}-\Phi^{\top})^{-1}$ is also known as the Ghosh inverse \cite{Ghosh58}, a supply-driven alternative to the more common demand-driven input/output (I/O) analysis \cite{Leontief36}. 
From the values of ${x,\Phi,q}$, other common MFA quantities of interest (QoIs) can be derived, such as mass flows for each connection ($z_{ij}=\phi_{ij}x_i$), and the sums, products, and ratios of mass flows. We represent these QoIs through a vector-valued function, $G(\Phi,q)$.
These QoIs (the components of $G$) typically correspond to the same quantities collected as MFA data (the components of $y$). 

We further refine our notation to introduce $\theta_m=\{\phi_{ij},q_i | \phi_{ij}\neq \text{const.},q_i\neq \text{const.} \,\text{under}\, M_m\}$ to collectively describe the set of all uncertain model parameters (i.e., of existing connections and external inflows) under a given network structure $M_m$.
For example, if the mass flow $z_{ij}$ is not present in the network structure $M_1$ but is in $M_2$, then $\phi_{ij}$ will always be zero and not be included in $\theta_{m=1}$, but $\phi_{ij}$ would be a non-trivial parameter and be part of $\theta_{m=2}$. Hence, what $\theta_m$ entails is dependent on the network structure $M_m$, and we emphasize this point with its subscript $m$.
This notation allows us to write succinctly $G(\Phi,q;M_m)=G(\theta_m;M_m)$.

\subsection{Parametric uncertainty under a fixed network structure}
\label{ss:parametric_uncertainty}

The parametric uncertainty, as represented by the parameter posterior distribution, under a fixed network structure $M_m$ can be obtained using Bayes' rule:
\begin{equation}
\label{e:bayes}
    p(\theta_m|y,M_m)=\frac{p(y|\theta_m,M_m)\,p(\theta_m|M_m)}{p(y|M_m)},
\end{equation}
where $p(\theta_m|y,M_m)$ is the posterior PDF, $p(y|\theta_m,M_m)$ is the PDF for the likelihood, $p(\theta_m|M_m)$ is the prior PDF, and $p(y|M_m)$ is the marginal likelihood (also known as model evidence).
We assign multi-variate Dirichlet prior distributions to the allocation fractions $\phi_{ij}$'s to automatically satisfy mass balance without needing to introduce additional constraints, while using independent truncated normal distributions with a non-negative lower bound for the mass flow input ($q_i$'s) priors. Different approaches can be adopted to assign hyper-parameters for the prior distributions (e.g., the concentration parameters of the Dirichlet, the mean and variance of the truncated normal). For example, Dong \textit{et al.} \cite{Dong23} used expert elicitation, via online surveys, to define informed priors that help to reduce the volume of data that must be collected to reach a desired reduction in uncertainty. Less taxing options include defining informed priors based on historical data or using non-informative priors \cite{Lupton18}.

The likelihood models the probability of obtaining the collected data, $y$, conditioned on the model parameters, $\theta_m$, and network structure $M_m$; thus, it provides a probabilistic measure on the mismatch between the observations, $y$, and the model prediction of the corresponding QoIs, $G(\theta_m;M_m)$. One option is to form the likelihood through an additive noise data model \cite{Dong23}. In this article, we adopt a relative noise model:
\begin{equation}
\label{e:relative_error}
    y_k=G_k(\theta_m;M_m)(1+\epsilon_k),
\end{equation}
where the subscript $k$ indicates the $k$th component of the vector, and $\epsilon_k\sim \mathcal{N}(0,\sigma_k^2)$ is also independent. Subsequently, we obtain
\begin{equation}
\label{e:likelihood}
   p(y|\theta_m,M_m)=
   \prod_{k=1}^{n_y}p_{\epsilon_k}\left(\frac{y_k}{G_k(\theta_m;M_m)}-1\right)=
   \prod_{k=1}^{n_y}\frac{1}{\sqrt{2\pi}\sigma_k}\exp\left[-\frac{\left(\frac{y_k}{G_k(\theta_m;M_m)}-1\right)^2}{2\sigma_k^2}\right]
\end{equation}
due to the independence of $\epsilon_k$'s.
Existing literature has explored two approaches to assigning values to the data noise standard deviation, $\sigma_k$. Lupton and Allwood \cite{Lupton18} assigned a fixed value (e.g., $\sigma_k=0.1$, corresponding to a level of $\pm 10\%$ relative noise) while Dong \textit{et al.} \cite{Dong23} modeled $\sigma_k$ as an unknown parameter and inferred it from data. They showed that with multiple years worth of data, the uncertainty in $\sigma_k$ can be significantly reduced; however, the computational cost to generate 10,000 posterior samples greatly increases from 3 hours to 17 hours for a network of 67 nodes and 169 flows informed by 95 data records, using a computer with Intel(R) CoreTM i7-11800H CPU, 2.30 GHz.

When considering candidate network structures, an observation (data record) may pertain to a flow or node that is deemed non-existent in some of the structures. In this scenario, the practitioner may either exclude the data record (as we recommend) or establish specific likelihood models to incorporate them (see S2). 

\subsection{Network structure uncertainty}
\label{ss:network_structure_uncertainty}

\subsubsection{Generating candidate network structures $M_m$}
\label{sss:model_generation}

To quantify the network structural uncertainty, the first step is to generate candidate node-and-flow structures for consideration. Here, we restrict the network structure uncertainty to the connectivity between nodes, excluding uncertainty on whether nodes exist. An intuitive approach is to generate a pool of candidate network structures by considering every possible permutation of connections between the nodes. However, such an exhaustive approach would lead to an exponential number of candidate structures, $2^{n_p^2}$, an infeasible consideration. Instead, we recommend using a combination of \textbf{exploitation} and \textbf{exploration} to create a sensible pool of candidate structures. Exploitation: A practitioner should extract existing MFA network structures from the literature (if they exist) and/or enlist domain experts to suggest and critique candidate network structures. Exploration: A practitioner may include a number of semi-randomized ``wild-guess'' network structures to help increase diversity. Once a total of $n_L$ connections (targeted connections) are identified, where $n_L$ is the combined number of network ``mutations'' from exploitation and the ``wild-guesses'', a total of $2^{n_L}$ network structures can be formed to generate the candidate pool from the complete permutation of the $n_L$ connections. We hypothesize that the combination of exploitation and exploration will enable an inclusive population of plausible candidate network structures, and allow us to find a better network structure than if using exploration or exploitation alone.

\subsubsection{Network structure prior $p(M_m)$}
\label{sss:model_prior}

Once the network structure candidates are established, we can either assign equal prior probabilities to all the candidates considered (i.e., a non-informative prior), or craft an informative prior. For example, an informative prior could be based on a complexity ranking to penalize more complex network structures \cite{Jeffreys98}, aligning with the principle of Occam's razor.
Alternatively, like the prior distributions for parametric uncertainty, expert elicitation can be applied, where a prior distribution is aggregated from individual experts, under the assumption that the existence of individual connections are independent from each other:
\begin{equation}
\label{e:model_prior}
    p(M_m)=\prod_{l=1}^{n_L} p_{\text{exist},l}^{d_{ml}}(1-p_{\text{exist},l})^{1-d_{ml}}
\end{equation}
where $p_{\text{exist},l}$ is, from individual experts, the aggregated probability of the $l$th connection existing in the network structure, and $d_{ml} \in \{0,1\}$ is an indicator variable associated with the state of the connection $l$ (i.e., $d_{ml}=1$ if the connection exists, 0 otherwise). Readers are directed to Dong \textit{et al.} \cite{Dong23} for details on aggregation of expert elicited priors. 

For the example model, we assign equal priors to both network structure candidates (\cref{f:toy_model} second row).

\subsubsection{Network structure posterior $p(M_m|y)$}
\label{sss:averaging}

\begin{figure}
    \centering
    \includegraphics[width=\textwidth,height=\textheight,keepaspectratio]{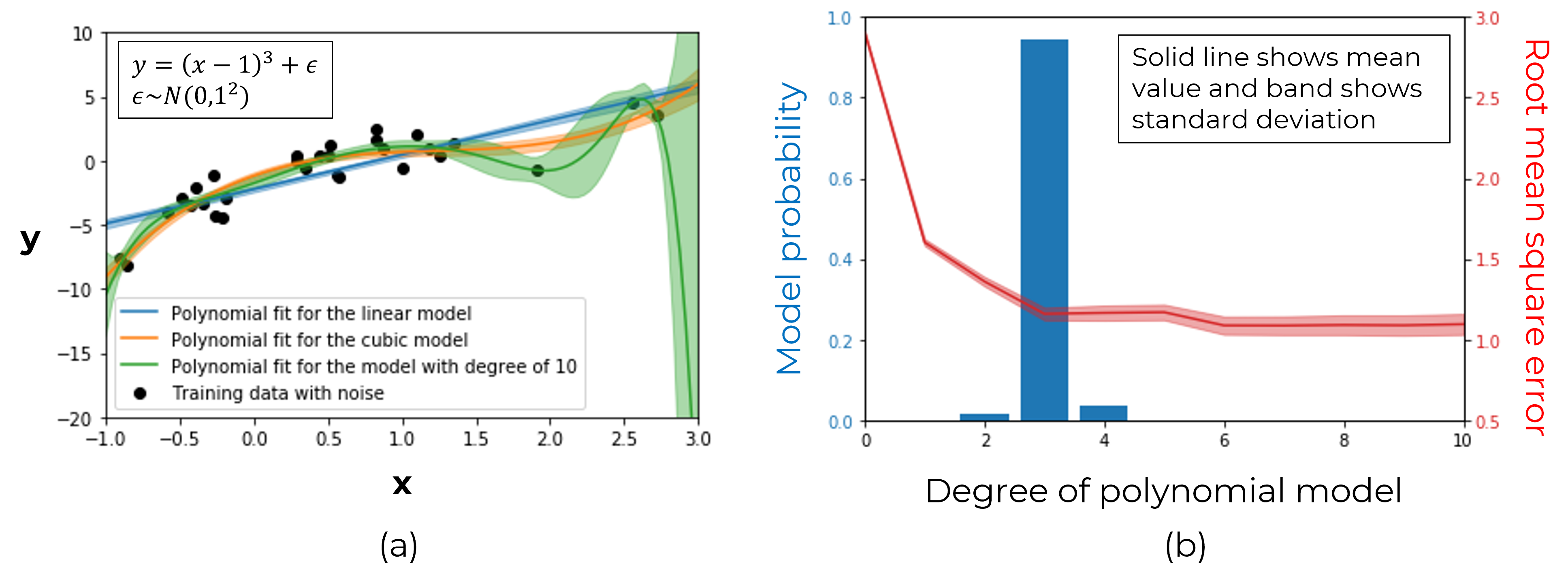}
    \caption{(a) Bayesian fitting of polynomial models to training data generated from a cubic function with noise $\epsilon$; solid lines show the mean fit and band shows one standard deviation from mean on each side.
    (b) Polynomial model posterior probabilities  following Bayesian model selection in \cref{e:bayes_model}, and the root mean squared error based on the individual polynomial models.
    }
    \label{f:regression}
\end{figure}

Two common model selection methods are based on comparing the models' Alkaike Information Criterion (AIC) \cite{Akaike74} and Bayesian Information Criterion (BIC) \cite{Schwarz78}. AIC and BIC are easy to compute and apply; however, they do not account for any prior knowledge on the network structures and do not provide a probabilistic measure for the network structure candidates. Instead, we follow the ideas of Bayesian model selection~\cite{Kass1995,Wasserman2000,Friel2012} and apply Bayes' rule to the network structure $M_m$, to arrive at its posterior probability distribution conditioned on the observed data $y$:
\begin{equation}
\label{e:bayes_model}
    p(M_m|y)=\frac{p(y|M_m)\,p(M_m)}{p(y)},
\end{equation}
where $p(M_m)$ is the network structure prior probability, $p(y)$ is the network structure marginal likelihood, and $p(y|M_m)$ is the PDF for the ``model likelihood'', which is the same term as the model evidence in the denominator of \cref{e:bayes}. In contrast to AIC or BIC, network structure posterior probability is more computationally expensive to evaluate due to the need to estimate $p(y|M_m)$. However, we still deem the Bayesian approach to model selection in \cref{e:bayes_model} favorable because, unlike AIC and BIC, it assigns all network structure candidates with a probability measure as justified by the observational evidence $y$. As a result,  the risk of losing an important true feature depicted by one of the those models is reduced, as a Bayesian averaged model (i.e., taking the posterior expectation following $p(M_m|y)$) will be a weighted average from all candidate models (see \cref{ss:mass_flow_uncertainty}). The model posterior does not always favor the most complex model, but instead strikes a balance between model simplicity and its ability to fit the data. We demonstrate this notion in \cref{f:regression}, which shows a cubic equation to be the more probable model for a series of noisy data (generated from a cubic function), despite higher order models resulting in a better fit to the data.

Using \cref{e:bayes_model}, we can also compare the relative probability of two models being the true network structure by taking the posterior ratio (PR):
\begin{equation}
\label{e:posterior_ratio}
    \text{PR}_{mn}=\frac{p(M_m|y)}{p(M_n|y)}
    =\frac{p(y|M_m)\,p(M_m)}{p(y|M_n)\,p(M_n)}=B_{mn}\, \frac{p(M_m)}{p(M_n)},
\end{equation}
where $B_{mn}$, commonly known as the Bayes factor, is the ``model likelihood ratio'' and indicates the model $M_m$ is B-times more likely to be the true network structure than the model $M_n$. Jeffrey \cite{Jeffreys35} provided guidelines for interpreting how strongly the Bayes factor indicates evidence for preferring one model over the other. Here, we apply the same guidelines to interpret the PR that also incorporates prior knowledge, see \cref{t:bf_guideline}.

In our work, we adopt the sequential Monte Carlo (SMC) algorithm~\cite{Chopin2020} implemented in PyMC3. SMC is a method that uses particles to characterize the parameter posterior $p(\theta_m|y,M_m)$ in \cref{e:bayes}, but also provides an estimate of the marginal likelihood $p(y|M_m)$ in its denominator as a by-product~\cite{Drovandi14}. This subsequently allows us to compute the PR terms in 
\cref{e:posterior_ratio}.

\begin{table}[]
    \centering
    \caption{Guidelines for interpreting how strongly the posterior ratio 
    provide evidence for the preference of one network structure over the other (based on guidelines from \cite{Jeffreys35}).}
    \begin{tabular}{l|c}
     \textbf{Quantitative result} & \textbf{Evidence in favor of $M_m$ being the better model is:} \\\hline
        $0.0<\log_{10} \text{PR}_{mn}\leq 0.5$ & Non-substantial \\
        $0.5<\log_{10} \text{PR}_{mn}\leq 1.0$ & Substantial \\
        $1.0<\log_{10} \text{PR}_{mn}\leq 1.5$ & Strong \\
        $1.5<\log_{10} \text{PR}_{mn} \leq 2.0$ & Very strong \\
        $2.0<\log_{10} \text{PR}_{mn}$ & Decisive
    \end{tabular}
    \label{t:bf_guideline}
\end{table}

\subsection{MFA posterior-predictive uncertainty}
\label{ss:mass_flow_uncertainty}

For a given network structure $M_m$, once the its parameter posterior is obtained, we can compute the posterior-predictive distribution for any QoIs in an MFA (e.g., mass flows) via
\begin{equation}
    p(G|y,M_m)=\int p(\theta_m|y,M_m) G(\theta_m;M_m) \, \text{d}\theta_m,
\end{equation}
where $G$ represents any MFA QoIs (including the connection mass flows $z_{ij}$'s and nodal mass flows $x_i$'s). 
Then, we can calculate the Bayesian model averaged posterior-predictive that now incorporates both parametric uncertainty and network structure uncertainty:
\begin{equation}
\label{e:mass_flow_weighted}
    p(G|y)=\mathbb{E}_{M_m|y}\left[\mathbb{E}_{\theta_m|y,M_m}\left[G(\theta_m; M_m)\right]\right] =  \sum_{m=1}^{n_m} p(M_m|y) p(G|y,M_m),
\end{equation}
where $n_m$ is the total (finite) number of candidate network structures. Essentially, $p(G|y)$ results from taking expectations over both the model posterior and the parameter posterior (i.e., averaging $G$ over the possible models and parameters weighted by their posterior probabilities). 
\Cref{e:mass_flow_weighted} can be used to derive a final Bayesian-averaged MFA model as shown for the example problem in \cref{f:toy_model} (bottom-left).

To estimate $p(G|y)$, we approximate the integral for $\theta_m$ using the posterior samples obtained from SMC. The $p(M_m|y)$ terms can be obtained from \cref{e:bayes_model} where the $p(y|M_m)$ terms are provided by SMC as a by-product and the denominator can then be easily computed via $p(y)=\sum_{m=1}^{n_m} p(y|M_m)p(M_m)$.

\subsection{Decision making for environmental sustainability under MFA uncertainty} 
\label{ss:decision_making}

The uncertainty of the mass flows in an MFA, calculated using Bayesian model averaging, can be translated into uncertainty in the environmental impacts (EI) of the associated system via:
\begin{equation}
\label{e:system_impact}
    \text{EI}_{\text{system}}=e^{\top} x,
\end{equation} 
where $e$ is a vector of environmental impact intensities (e.g., kg.CO\textsubscript{2eq.}/kg\textsubscript{throughput}) for each node and $x$ is the corresponding vector of nodal mass flows. Uncertainty regarding $e$ may also be included at this stage. When exploring options for improving the environmental sustainability of a supply chain, it is often helpful to attribute a system's environmental impacts to the final demand (consumption) sectors that drive the supply chain. These sectors are typically represented as the terminal nodes in an MFA. If no additional terminal loss nodes are present then, using classic demand-driven I/O analysis \cite{Leontief86,Kitzes13}, the uncertainty of the mass flows in an MFA can be translated into uncertainty of the environmental impact intensity (EII, per unit of consumption) attributable to node $i$, via.
\begin{equation}
\label{e:cumulative_emission}
    \text{EII}_{i}=e^{\top} L_{\cdot i}
\end{equation}
where $L=(\mathbb{I}-A)^{-1}$ is the Leontief inverse with $L_{\cdot i}$ indicating its $i$th column and $A$ being the technical coefficient matrix that describes the input mass flow to a given node from every other node and calculated from the nodal mass flows and allocation fractions:
\begin{equation}
\label{e:A_matrix}
    A_{ij}=\frac{z_{ij}}{x_j}=\frac{\phi_{ij}x_i}{x_j}.
\end{equation}
Many readers will be familiar with \cref{e:cumulative_emission} which is why we present it here; however, we also derive in S3 
an equivalent method for calculating the system-wide emissions attributable to a consumption node using the allocation fraction matrix $\Phi$, avoiding the need to calculate the Leontief inverse and offering a computational advantage.
Using the I/O method, \cref{f:toy_model} (bottom right) shows the PDFs representing the emission-intensity of the three consumption nodes in the example problem. The EII of a consumption node is equal to the environmental impact savings if mass flows into the same consumption node are decreased by one unit of mass (assuming a linear relationship between consumption and impacts). 

In supply chain MFAs, besides final demand sectors, there can also be a terminal loss node; e.g., from oxidation in liquid metal processing. As the generation of losses is not the motivation for the supply chains' existence, it is desirable to reallocate the environmental impacts attributable to the terminal loss node to other terminal (consumption) nodes. This reassignment can be achieved by considering a perturbation in demand from a consumption node and calculating the effect on emissions attributable to the consumption and loss nodes when the nodal process yield loss fractions remain constant (as is likely in reality). First, the vector of terminal nodal demand ($F$) can be rearranged into a vector containing the consumption demand ($F_{\text{cons}}$) and a vector containing the loss node throughput re-attributed to the loss generation processes ($F_{\text{loss}}$). Consider the example problem in \cref{f:toy_model}, if node 5 serves as a loss node rather than a consumption node, the loss vector $F_{\text{loss}}$ for this problem would have non-zero elements at $F_{\text{loss},3}$ and $F_{\text{loss},8}$, and the consumption vector would have a zero element at $F_{\text{cons},5}=0$. The new loss node throughput ($F_{\text{loss,new}}$) from any changes to the baseline consumption demand ($F_{\text{cons,base}}$) can be expressed as \cref{e:new_node_eqn1}, where $\Gamma$ is a diagonal matrix of nodal yield losses (0--1) ($\text{Diag}\{\Gamma\}=\Phi_{\cdot \text{loss}}$ extractable from the column of allocation fraction matrix $\Phi$ corresponding to the loss node), and $x_{\text{new}}$ is the vector of new nodal mass flows induced by the change in terminal nodal demand. The new vector of terminal nodal demand ($F_{\text{new}}$) can be expressed as \cref{e:new_node_eqn2}: the addition of the new consumption demand ($F_{\text{cons,new}}$) and $F_{\text{loss,new}}$. Note that the $F_{\text{new}}$ vector contains zero-demand at the terminal loss node element.
Finally, using classic I/O analysis, the new vector of nodal mass flows ($x_{\text{new}}$) is expressed as \cref{e:new_node_eqn3}: the product of the unchanging Leontief matrix and $F_{\text{new}}$. These relationships are summarized as:
\begin{align}
        F_{\text{loss,new}}&=\Gamma x_{\text{new}},\label{e:new_node_eqn1}
\\
        F_{\text{new}}&=F_{\text{cons,new}}+F_{\text{loss,new}},\label{e:new_node_eqn2}
\\
        x_{\text{new}}&=LF_{\text{new}}\label{e:new_node_eqn3}
.
\end{align}
By solving \cref{e:new_node_eqn1,e:new_node_eqn2,e:new_node_eqn3}, the new nodal mass flows after a perturbation in consumption demand, assuming the nodal yield loss fractions remain unchanged, is
\begin{equation}
\label{e:node_rectified}
    x_{\text{new}}=(\mathbb{I}-L\Gamma)^{-1}LF_{\text{cons,new}}.
\end{equation}
The EII attributable to node $i$, rectified to reallocate the environmental impacts otherwise attributable to the loss node to consumption node $i$, is therefore the difference between the EI of the associated system before and after a change in consumption demand at node $i$, divided by the change in consumption demand at node $i$:
\begin{equation}
\label{e:emission_rectified}
    \textbf{Rectified }\text{EII}_i=\frac{\text{EI}_{\text{base}}-\text{EI}_{\text{new}}}{F_{\text{cons,base},i}-F_{\text{cons,new},i}}
\end{equation}

Having derived the emission intensity distributions for each consumption node, the next step is to determine which node should be prioritized for consumption reduction efforts. Decisions are often made by weighing criteria such as the potential benefits, risks, and the certainty of the outcome \cite{McPhail18}. When the benefit of choosing a decision option is expressed as a PDF, then these criteria can be translated into quantitative metrics to aid decision-making. McPhail \textit{et al.} \cite{McPhail18} reviewed multiple popular decision-making metrics. By analyzing the benefit distribution associated with each option, a decision can be made, for example, to maximize: i. the mean benefit (maximizing the expected benefit); ii. the inverse of the coefficient of variation  (maximizing the certainty of the benefit); iii. the maximum value from the distribution (``maximax'': maximizing the best possible outcome); or, iv. the minimum value from the distribution (``maximin'': maximizing the benefits even under the worst outcome). The choice of decision-making metric depends on the risk-tolerance of the decision-maker. Note that as the tails of normal distributions extend to infinity, then it is often necessary to convert the the maximax and maximin criteria into maximizing a high (e.g., 95th) and low (e.g., 5th) percentile criterion respectively. Applying these metrics to the example problem, it can be observed in \cref{f:toy_model} (bottom) that consumption reduction efforts should focus on node 4 to maximize both the expected emission reduction and the emission reduction under a best possible outcome; whereas, consumption reduction efforts should focus on node 5 to maximize both the certainty of the expected emission reduction and the emission reduction under a worst possible outcome. In contrast, reducing the consumption at node 9 is not the priority under any of the decision-making metrics analyzed.
\section{Case study on the U.S. steel flow}
\label{s:USGS}

We demonstrate the use of Bayesian inference to incorporate network structure uncertainty in MFA through a case study on the U.S. annual flow of steel in 2012. This year is chosen for the sake of consistency with earlier work on expert elicitation and data nose learning in MFA using Bayesian inference \cite{Dong23}; however, any year could have been used. The mass flow uncertainty results are used to inform a decarbonization strategy based on reducing demand for steel in different end-use sectors. All data and code used in this case study are available online (see the SI).

\subsection{Constructing candidate network structures}
\label{ss:cs_model}
We first extract a baseline network structure from Zhu \textit{et al.}'s \cite{Zhu19} study on the 2014 U.S. steel flow. Their MFA includes 270 metal flows connecting 55 nodes. Following the exploitation and exploration methodology described in \cref{sss:model_generation}, we generate additional candidate network structures by identifying four targeted connections from the map.

\textbf{Exploitation}: The baseline network structure was discussed with industry experts from Nucor and U.S. Steel and compared to steel MFAs for other years \cite{Dong23} and geographies \cite{Cullen12}. This revealed a targeted connection from post-consumer steel scrap to the blast furnace (BF, connection index 1). This connection is absent in the baseline structure from Zhu \textit{et al.} and questioned by the industry experts; however, it is present in Dong \textit{et al.} \cite{Dong23} and in the United States Geological Survey (USGS) steel statistics \cite{USGS12a}.

\textbf{Exploration}: To help increase the diversity of the candidate network structure pool, we consider the presence or absence of connections between other nodes: the flow between scrap and the basic oxygen furnace (BOF) (connection index 2), between BOF continuously cast slab and the rod and bar mill (connection index 3), and between BOF continuously cast slab and the section mill (connection index 4). These connections are highlighted in the Sankey diagram in \cref{f:sankey diagram}. The existence of a scrap flow into the BOF (index 2) is well-known and acts in this case study as a basic test of the model selection methodology. It is also well-known that BOF continuous casting in the U.S. is used for high-quality, flat-product production (i.e., connected to the hot strip mill and plate mill); however, it is less certain whether there are connections to the rod and bar (index 3), and section mills (index 4).

A pool of $2^4=16$ candidate network structures are generated from the complete permutation of the 4 targeted connections. The network structures are described using a 4-digit binary code to indicate whether the indexed connection is present (1) or absent (0). For example, all 4 targeted connections exist in the network structure, 1111; whereas, out of the 4 targeted connections, only the flow between scrap and the BOF (index 2) exists in the network structure, 0100.

\subsection{Constructing prior distributions for network structure and parameters}
\subsubsection{Network structure prior}
A panel of three academic experts on steel industry sustainability from the University of Michigan was interviewed to assess the probability of existence, based on their expert judgment, of each targeted connection identified in \cref{ss:cs_model}. The panel consensus was a 10\% probability for the flow from post-consumer steel scrap to the blast furnace (index: 1), a 95\% probability for the flow from scrap to the BOF (index: 2), a 15\% probability for the flow from BOF continuous casting to the rod and bar mill (index: 3), and a 15\% probability for the flow from BOF continuous casting to the section mill (index: 4). The expert elicited prior probability for each of the 16 network structures was then derived based on  \cref{e:model_prior} and is shown in the bar chart in \cref{f:bf_table}.

\subsubsection{Parameter priors}
To form informative parameter priors, results from expert elicitation are used for upstream allocation fractions ($\phi_{ij}$'s) and external inputs ($q_i$'s), while non-informative priors are used for the downstream allocation fractions. Readers are directed to Dong \text{et al.} \cite{Dong23} for the details of the expert elicitation and prior aggregation method, with the resulting prior distributions provided in the S4. 

\subsection{U.S. steel flow and emissions data collection}
\label{ss:cs_data}

Steel flow data were collected from the USGS \cite{USGS12a, USGS12b, USGS12c}, World Steel Association (WSA) \cite{WSA12} and Zhu \textit{et al.} \cite{Zhu19}. A complete record of all collected MFA data is provided in S5. 
For the likelihood, a fixed relative data noise level with $\sigma_k=10\%$ is applied to contain computational cost, which has been shown to be a reasonable for this dataset based on the noise-learning results from Dong \textit{et al.} \cite{Dong23}.

Estimated nodal emissions intensities for each U.S. process in the steel network are shown in S7. This analysis focuses on domestic emissions and no impacts are assigned to import nodes.

\subsection{Case study results and discussion}
\label{ss:uncertainty_result}

For each network structure (containing approximately 180 parameters) it takes about 5 hours using an Intel(R) Core i7-9700K CPU, 3.60GHz to generate 10,000 posterior samples using the SMC implementation of PyMC3. This translates to a total computational time of around 80 hours across the 16 candidate network structures. While significant, the computational time could be reduced by, for example, using multi-core processors to run algorithms that can be parallelized or applying approximate Bayesian techniques such as variational inference \cite{Dong23,Blei17}.

\subsubsection{Network structure and mass flow uncertainty}

\begin{figure}
    \centering
    \includegraphics[width=\textwidth,height=\textheight,keepaspectratio]{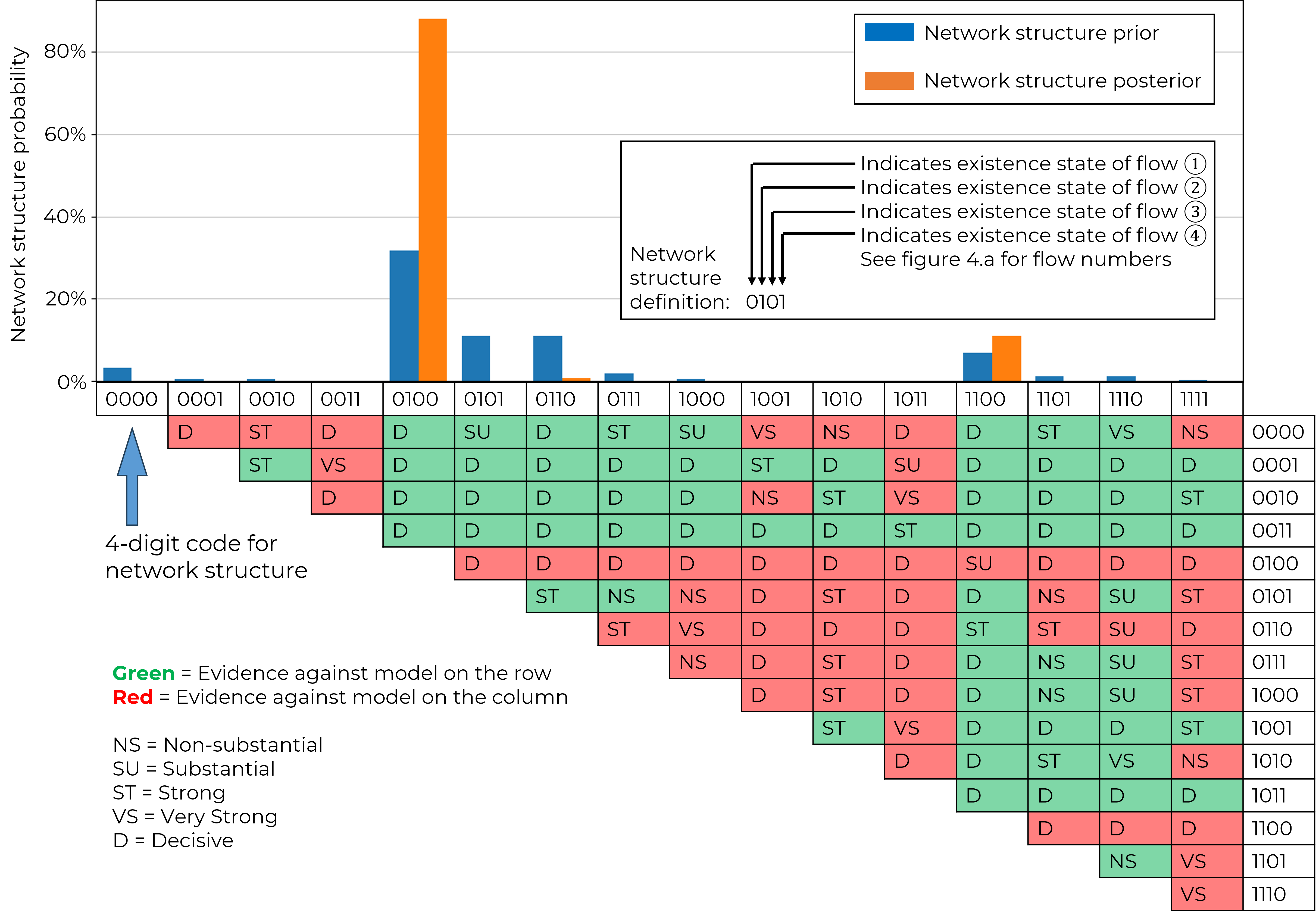}
    \caption{Posterior probability and the pair-wise posterior ratio interpretation for U.S. steel flow network structure candidates.
    }
    \label{f:bf_table}
\end{figure}

The posterior probability for each of the 16 candidate network structures is shown in \cref{f:bf_table} along with their pairwise comparisons using the posterior ratios. \Cref{f:bf_table} shows that network structure 0100 (the original structure from Zhu \textit{et al.}) achieves the highest posterior probability at 88\%, increased from a prior probability of 32\%. Conversely, the probabilities for the network structures where a flow exists from continuously cast slab to either the rod and bar mill or section mill decrease from the prior to the posterior. The pairwise comparison indicates that evidence supporting 0100 is ``Decisive'' against all other network structures, except 1100 which the evidence remains ``Substantial''. Furthermore, there is ``Decisive'' evidence against network structures where scrap flowing into the BOF is absent (e.g., 1011).

\begin{figure}
    \centering
    \includegraphics[width=\textwidth,height=\textheight,keepaspectratio]{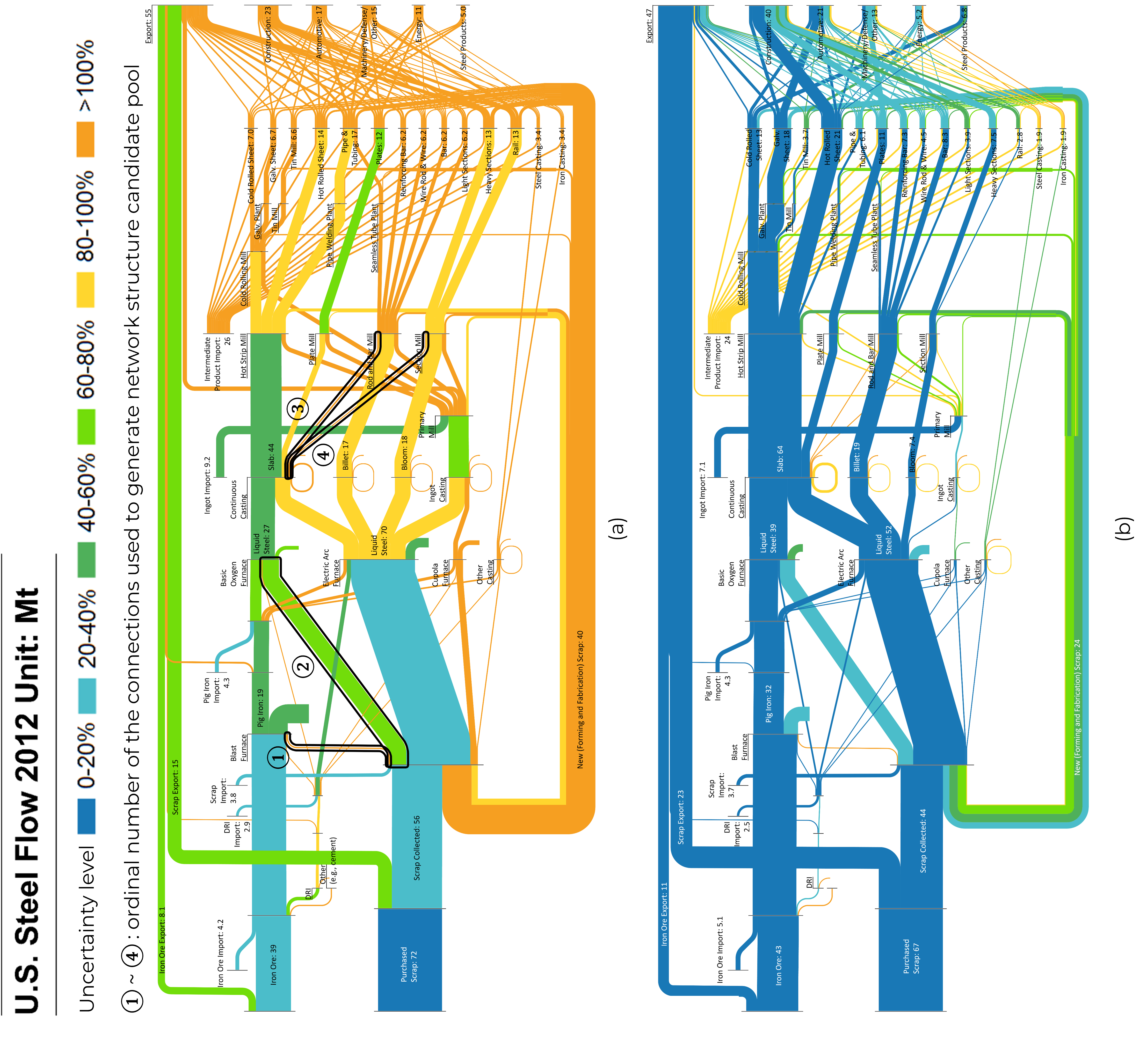}
    \caption{Bayesian model averaged (a) prior- and (b) posterior-predictive mass flows for the U.S. steel flow in 2012.
    All numbers on the flows refer to the mean of the mass flow in units of million metric tons (Mt). The uncertainty percentages refer to the flow standard deviation as a percentage of the mean of the mass flow. All mass flows refer to steel except for the iron ore flows that include the non-iron mass (e.g., oxygen and gangue).
    }
    \label{f:sankey diagram}
\end{figure}

\Cref{f:sankey diagram} presents the Bayesian model averaged prior and posterior mass flows as Sankey diagrams for the U.S. steel flow map in 2012. The width and color of the lines indicate the size and uncertainty of the flow, with a darker blue indicating a lower uncertainty level. As an averaged model, the network structure in \cref{f:sankey diagram} includes all targeted connections (i.e., 1111). However, given the high posterior probabilities of network structures 0100 and 1100, the mass flows in the posterior Sankey diagram in \cref{f:sankey diagram}.b largely reflect the mass flows in the posterior Sankey diagrams for these two structures with the fourteen other structures having only minor contributions. The prior and posterior Sankey diagrams for each of the 16 individual candidate network structures can be found in S6. 

\subsubsection{Informed decision making for decarbonization via demand reduction}
\label{ss:dm_result}

\begin{figure}[h]
    \centering
    \includegraphics[width=\textwidth,keepaspectratio]{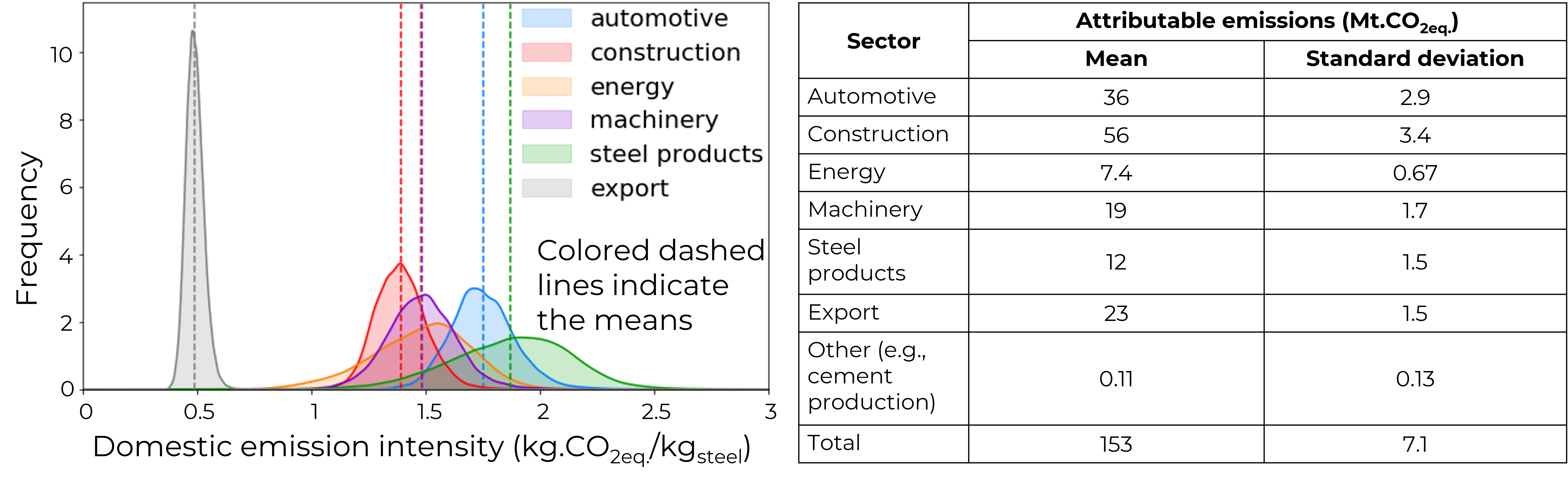}
    \caption{Domestic emissions and emission intensities attributable to U.S. consumption sectors plus export.}
    \label{f:emission_curve}
\end{figure}

The nodal emission intensities ($e$, from S7) and the nodal mass flows ($x$, from the posterior Bayesian model averaged MFA) are used in \cref{e:system_impact} to calculate the total domestic emissions attributable to the 2012 U.S. steel system: a mean of 153 Mt.CO\textsubscript{2eq.} and a standard deviation of 7.1 Mt.CO\textsubscript{2eq.}. \Cref{f:emission_curve} shows the attributable emissions and emissions intensities for the U.S. consumption sectors plus export, calculated using \cref{e:emission_rectified}. \Cref{f:emission_curve} shows that the automotive and steel product consumption sectors have the highest mean emission intensities. This is because these sectors use significant quantities of high-quality sheet metal, much of which is produced using the emission-intensive BF-BOF primary steelmaking route due to sheet metal's very low tolerance to copper which is abundant is post-consumer scrap \cite{Daehn17,WEF23}. Export has the lowest mean emission-intensity as this sector is dominated by export of iron ore and post-consumer scrap that has yet to undergo emission-intensive processing into steel semi-finished products.

Using the decision making criteria introduced in \cref{ss:decision_making}, \cref{t:emission_criteria} shows a prioritization for reducing steel demand across the consumption sectors (plus export) dependent on the decision maker's appetite for risk. \Cref{t:emission_criteria} shows that under circumstances in which the decision maker wishes to maximize the savings under the worst outcome (``Maximize low savings''), then demand reduction efforts should focus on the automotive sector. Furthermore, if the decision maker is wishing to maximize the certainty of the expected savings, then demand reduction efforts should focus on the ``Other'' sector. Finally, if the decision maker is wishing to either maximize the expected savings or savings under the best outcome (``Maximize high savings''), then demand reduction efforts should focus on the steel products sector. This result is reflected in the distributions shown in \cref{f:emission_curve}, where the right-hand tail of the steel products distribution extends further than for the other sectors, indicating higher emission savings per unit of reduced consumption under the best outcome. Limitations to using the emissions intensity per unit of consumption to prioritize consumption reduction efforts include that the I/O analyses used to derive the emissions intensities are based on linear models that assume a constant, fixed ratio of inputs are used to produce a sector's output \cite{Kitzes13}. Focusing on emissions intensity per unit of consumption also ignores the overall scope for change; e.g., while the construction sector is not the most emissions intensive consumption sector per unit consumed, it accounts for more emissions (and steel produced) than any other single sector (see \cref{f:emission_curve}); therefore, the overall scope for reducing demand may be largest in the construction sector. Relatedly, focusing on emissions intensity does not account for the difficulty of implementing demand reduction efforts across different consumption sectors, even though this would be a crucial factor in the decision-making process.

\begin{table}[]
    \centering
    \caption{Decision criteria for decarbonization (``savings'') through demand reduction at U.S. steel consumption sectors.
    } \includegraphics[width=\textwidth,keepaspectratio]{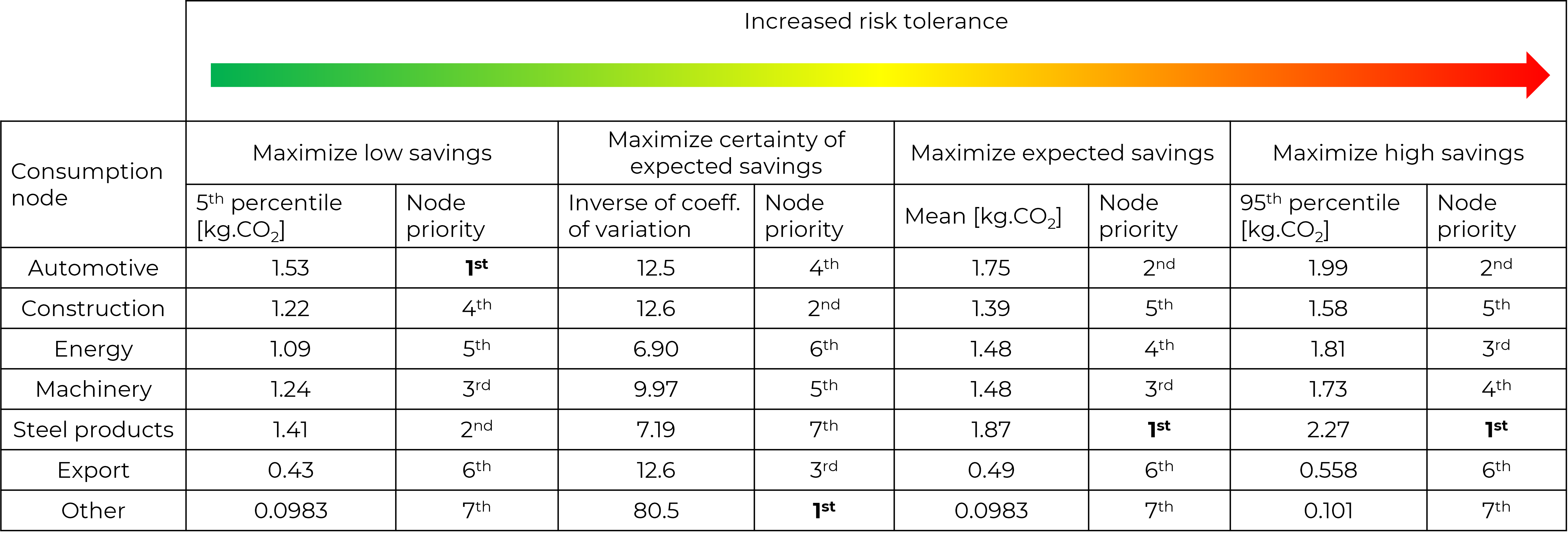}
    \label{t:emission_criteria}
\end{table}
\section{Conclusions and future work}
\label{s:discussion}

The Bayesian framework provides a systematic and mathematically rigorous method for incorporating network structure uncertainty into MFA uncertainty results. Comparing the posterior ratios of different network structures allows a practitioner to determine the level of evidence in favor of one model structure versus another, and Bayesian model averaging allows the practitioner to gain insights from all the candidate models. The value of rigorous uncertainty quantification is to enable more informed decision making. The holistic MFA mass flow uncertainty results generated by Bayesian model averaging can be readily combined with the I/O method to help prioritize consumption reduction efforts. Through a case study on the 2012 U.S. steel flow, we demonstrate the expanded Bayesian framework for MFA and its utility in allowing more informed decision-making: the automotive industry is identified as a priority for demand reduction efforts that maximize the expected emissions savings. The code for this case study has been made available (via the SI) to help readers apply these methods.

Integrating network structure uncertainty into the uncertainty quantification in MFA may necessitate the collection of additional MFA data to reduce mass flow uncertainty to an acceptable level. This could be problematic as data collection is often a bottleneck in constructing comprehensive MFAs. Future work could address this challenge by combining the Bayesian framework with optimal experimental design~\cite{Chaloner95,Muller05,Ryan2016,Rainforth2023,Huan2024,Liao25} to prioritize data collection that most effectively reduces both network structure and mass flow uncertainty.


\section{Acknowledgments} 
The authors would like to thank all the steel experts interviewed for this study. This material is based upon work supported by the National Science Foundation under Grant No. \#2040013.



\section{Supporting information}

This Supporting Information (SI) document includes data, literature reviews and Sankey diagrams for individual candidate network structures helpful to understanding the main article as well as links to Python Scripts and collected MFA data used to conduct the case study.

Please go to this link (\url{http://remade.engin.umich.edu/MFA_NSF.htm}) for downloads of the following: 
\begin{itemize}
    \item presentation of the underlying data used to construct the Sankey diagrams in the main paper (Figure 4) in a numerical, tabular format;
    
    \item a Python script for performing both the Bayesian inference for both the parametric and network structure uncertainty using specified prior PDFs and collected MFA data; and

    \item a Python script for performing MFA Bayesian model averaging and decision-making using the rectified input/output (I/O) analysis.
\end{itemize}

\subsection{Different forms of data available for conducting MFAs}
\label{SI:data_form}

\begin{center}
\begin{longtable}{  p{4cm}|p{12cm}  } 
    \caption{Typical data forms available for conducting MFAs. Adapted from Kopec \textit{et al.} \cite{Kopec16}. Data is generally sparse, e.g., it is uncommon to have more than 1,000 data records for constructing a static supply chain MFA.}\\
    Data type & Example from U.S. steel flow MFA\\
    \hline
    Stated existence of a node or a flow between 2 nodes  &  USGS \cite{USGS14} reports a 24 Mt flow of pig iron between the blast furnace (BF) and basic oxygen furnace (BOF), revealing the existence of both nodes and the presence of a flow between them.\\
    Flow between 2 nodes & USGS \cite{USGS14} reports 24 Mt flow between the ``BF'' node and ``BOF'' node.\\
    Sums of flows & USGS \cite{USGS16} reports the sum of all continuous casting product flows (slabs, billets and blooms) at 88 Mt in 2016.\\
    Percentages of sums & AISI \cite{AISI15} reports construction sector taking up 40\% of total steel demands.\\
    Percentages to a destination & Omar \cite{Omar11} reports the average process yield when making an irregular sheet metal car side body panel as 38\%.\\
    Percentages from an origin & USGS \cite{USGS16} reports 70\% of the continuous casting products being slabs.\\
    Additional linear relations & WSA \cite{WSA09} reports oxidation losses from direct reduction and the blast furnace are equal\\
    Sequential multiplications & Milford \textit{et al.} \cite{Milford11} report blanking and stamping process yields separately, that can be multiplied together to get an overall sheet metal fabrication yield.
    \label{t:SI_data_type}
\end{longtable}
\end{center}

\subsection{Alternative method for likelihood modeling of targeted connections}
\label{SI:alt_likelihood}

When considering candidate network structures, an observation (data record) may pertain to a flow or node that is deemed non-existent in some of the structures. In this scenario, the practitioner may either exclude the data record (as we recommend) or establish specific likelihood models to incorporate them.

The inclusion of data records on the targeted connections that are missing in certain candidate network structures requires special treatment. As these flows are not present in the network structure, the observations should have no impact on the parametric posteriors from the inference regardless of the value of the observation:
\begin{equation}
\label{e:const}
    p(y_k|\theta_m;M_m)=C_k \quad \text{for} \quad \theta_m\in\text{supp}(\theta_m),
\end{equation}
where $C_k$ is a constant regardless of the value of $y_k$, and $k$ is the index for any observation on the missing connections. The choice of the constant value $C$ would impact the marginal likelihood (see Equation (4) in the manuscript) and therefore the fairness of Bayesian model selection. For a proper uniform distribution model, a compact support for the likelihood is required, and should be consistent across all candidate network structures. As a result, instead of a normal distribution for the parametric likelihood model (see Equation (6)), a truncated normal with properly selected upper and lower bounds $b_u$ and $b_l$ should be applied. As a result, the likelihood can be modified as follows:
\begin{equation}
\label{e:SI_missing_feature}
        p(y|\theta_m;M_m)= 
    \begin{cases}
        \displaystyle \prod_{l=1}^{n_y-n_L} p(y_l|\theta_m;M_m)\prod_{k=1}^{n_L} C_k=\prod_{l=1}^{n_y-n_L} p(y_l|\theta_m;M_m)\prod_{k=1}^{n_L} \frac{1}{b_{u,k}-b_{l,k}}, & \text{for}\,\, \theta_m\in\text{supp}(\theta_m)\\
        0,                 & \text{else}
    \end{cases},
\end{equation}
where $n_y$ is the total number of observations and $n_L$ is the number of data records on the targeted connections.

\subsection{Alternative method for a supply-driven Input/Output analysis}
\label{SI:ghosh_IO}

The alternative supply-driven I/O method to calculate the environmental impacts (EI) of the associated system avoids the calculation of the Leontief inverse \cite{Leontief36}, which provides computational benefits over the classic I/O method.
For any system, an ``emission-balance'' can be established at a given node $i$ where the total emissions attributable to the output nodal mass flows is equal to the sum of total emissions attributable to the input nodal mass flows and any emissions produced/captured at the process, which can be either due to emission released at the process or additional supplies at the node:
\begin{equation}
\label{e:SI_emission_balance}
    \text{EI}_{i}=\sum_{j=1}^{n_p} \text{EI}_j\cdot \phi_{ji} +e_{0,i}\cdot x_i+e_{q,i}\cdot q_i,
\end{equation}
where $\phi_{ji}$ is the allocation fraction, $e_0$ is a vector of emission-intensities produced/captured at each node, and $e_q$ is a vector of emission-intensity associated with the material inflow $q$.
Subsequently, \cref{e:SI_emission_balance} can be assembled into a matrix form:
\begin{equation}
\label{e:SI_emission_matrix_form}
    \text{EI}=(\mathbb{I}-\Phi^{\top})^{-1}(\text{EI}_0+\text{EI}_q),
\end{equation}
where $\text{EI}_0$ \text{ and } $\text{EI}_q$ are column vectors of the emissions produced/captured at each node and the embodied emissions of the material inflow $q$, respectively, with elements $\text{EI}_{0,i}=e_{0,i} x_i$ and $\text{EI}_{q,i}=e_{q,i} q_i$. Therefore, the environmental impact intensity (EII) at node $i$ is:
\begin{equation}
\label{e:SI_EII}
    \text{EII}_i=\frac{\text{EI}_i}{x_i}.
\end{equation}

\subsection{Parameter priors for the case study}
\label{SI:cs_prior}

In this section, we include the details of the informative priors ($\phi$ and $q$) used for the case study on the 2012 U.S. steel flow. The informative priors for this case study are obtained from expert elicitation through interviews with domain experts. Readers are directed to Dong \textit{et al.} \cite{Dong23} for the details of the methodology to conduct expert elicitation and prior aggregation from multiple experts. The elicitation process conducted is under the assumption of the network structure with all 4 targeted connections existent. As informative priors are applied to both the flows originating from scrap node and continuous cast slab node, adjustment is required to apply these informative priors to candidate network structures where one or more targeted connections are not present in the model. We delete the hyper-parameter(s) of the Dirichlet priors for allocation fractions corresponding to the targeted connection(s) if they do not exist in the candidate network structure, while keeping the rest of the hyper-parameters fixed. For example, the informative prior distribution used for the allocation fractions originating from continuous cast slab to hot strip mill, plate mill, rod and bar mill and section mill when all four connections are present is $\phi\sim Dir(11.46,2.11,2.82,1.81)$. In the case where the connection to the rod and bar mill does not exist, the revised informative prior for this set of allocation fraction will be $\phi\sim Dir(11.46,2.11,1.81)$.

For the exact values of the hyper-parameters used for informative priors, please see the inference code from the link below (\url{http://remade.engin.umich.edu/MFA_NSF.htm}).

\subsection{Case study: U.S. steel flow MFA collected data}
\label{SI:MFA_data}

\begin{center}
\begin{longtable}{  m{17em} m{3.5cm} m{3cm} m{2cm}  } 
  \caption{MFA data from 2012.}\\
  \hline
  \textbf{Description} & \textbf{Type} &\textbf{Value (Mt)} & \textbf{Source} \\ 
  \hline\hline
  Import to Iron Ore Consumption & External Input &  5.16 & 1 \\ 
  \hline
  Iron Ore Production & External Input & 54.7 & 1 \\ 
  \hline
  Iron Ore Production to Export & Flow & 11.2  & 1 \\ 
  \hline
  Iron Ore Consumption to Blast Furnace &  Flow  & 46.3  & 1 \\ 
  \hline
  Blast Furnace to Pig Iron &  Flow  & 32.1  & 3 \\
   \hline
  Import to DRI Consumption & External Input & 2.47 & 2 \\ 
  \hline
  DRI to Export &  Flow  &  0.01 & 2 \\ 
  \hline
  DRI Consumption to Blast Furnace &  Flow  & 0.049  & 2 \\ 
  \hline
  DRI Consumption to Basic Oxygen Furnace &  Flow   &  1.91 & 2 \\ 
  \hline
  DRI Consumption to Electric Arc Furnace &  Flow  &   1.62 &  2\\ 
  \hline
  DRI Consumption to Cupola Furnace &  Flow  & 0.01  & 2 \\ 
  \hline
  DRI Consumption to Other &  Flow  & 0.01  & 2 \\ 
  \hline
  Import to Pig Iron Consumption & External Input &  4.27 & 2 \\
  \hline
  Pig Iron to Export &  Flow  & 0.021  & 2\\ 
  \hline
  Pig Iron to Basic Oxygen Furnace &  Flow  &  31.5  & 2 \\ 
  \hline
  Pig Iron to Electric Arc Furnace &  Flow  &  5.79 & 2 \\ 
  \hline
  Pig Iron to Cupola Furnace &  Flow   &  0.057 & 2 \\ 
  \hline
  Pig Iron to Other &  Flow  & 0.046  &  2\\ 
  \hline
  Import to Scrap Consumption & External Input & 3.72 & 2 \\
  \hline
  Purchased Scrap to Scrap Collected & External Input & 70.98 & 2\\
  \hline
  Scrap Collected to Export &  Flow  &  21.4  &  2\\ 
  \hline
  Scrap Consumption to Electric Arc Furnace &  Flow   & 50.9  & 2 \\
  \hline
  Scrap Consumption to Cupola Furnace &  Flow  &  1.11 & 2 \\ 
  \hline
  Scrap Consumption to Other &  Flow  &  0.167 & 2 \\ 
  \hline
  BOF\_CC to Continuous Casting &   Flow & 36.281  & 4\\
  \hline
  HSM\_Yield to Hot Rolled Sheet &  Flow  & 19.544   & 3 \\ 
  \hline
  CRM\_Yield to Cold Rolled Sheet &   Flow  & 11.079  &  3\\ 
  \hline
  Plate Mill to Plates &  Flow  &  9.12 & 3 \\ 
   \hline
  RBM\_Yield to Reinforcing Bars &  Flow  & 5.65  & 3 \\ 
  \hline
  RBM\_Yield to Bars &  Flow  &  6.7 & 3\\ 
   \hline
  RBM\_Yield to Wire and Wire Rods &  Flow  & 2.784  & 3 \\ 
  \hline
  RBM\_Yield to Light Section &  Flow  & 2.13  &  3\\ 
 \hline
  SM\_Yield to Heavy Section &  Flow  &  5.03 &  3\\ 
  \hline
  SM\_Yield to Rail and Rail Accessories &  Flow  & 1.009  & 3 \\
  \hline
  PM\_Yield to Export &  Flow  & 0.817  & 3 \\
  \hline
  Tin Mill to Tin Mill Products &  Flow  & 2.009  &  3\\
  \hline
  Galvanized Plant to Galvanized Sheet &  Flow  &  16.749 &  3\\
  \hline
  Pipe Welding Plant to Pipe and Tubing &  Flow  & 2.165  & 3\\
  \hline
  Seamless Tube Plant to Pipe and Tubing &  Flow  & 2.162 & 3 \\
  \hline
  Electric Arc Furnace to Billet &  Ratio  & 0.333 & 5 \\
  \hline
  Electric Arc Furnace to Bloom &  Ratio  & 0.157 & 5 \\
  \hline
  Electric Arc Furnace to Ingot Casting &  Ratio  & 0.02 & 5 \\
  \hline
  Cold Rolled Sheet to Automotive & Ratio  & 0.25  & 5\\
  \hline
  Cold Rolled Sheet to Machinery & Ratio  & 0.079  & 5\\
  \hline
  Cold Rolled Sheet to Steel Products & Ratio & 0.313  & 5\\
  \hline
  Cold Rolled Sheet to Export & Ratio & 0.112 & 5\\
  \hline
  Galvanized Sheet to Construction & Ratio & 0.19  & 5\\
  \hline
  Galvanized Sheet to Automotive & Ratio & 0.42  & 5\\
  \hline
  Galvanized Sheet to Export  & Ratio  &  0.15 & 5\\
  \hline
  Hot Rolled Sheet to Construction & Ratio & 0.59  & 5\\
  \hline
  Hot Rolled Sheet to Automotive & Ratio & 0.133  & 5\\
  \hline
  Hot Rolled Sheet to Machinery  & Ratio &  0.108  & 5\\
  \hline
  Hot Rolled Sheet to Energy & Ratio & 0.01  & 5\\
  \hline
  Hot Rolled Sheet to Steel Products & Ratio & 0.0027  & 5\\
  \hline
  Hot Rolled Sheet to Export  &  Ratio & 0.065  & 5 \\
  \hline
  Pipe and Tubing to Construction & Ratio & 0.227  & 5\\
  \hline
  Pipe and Tubing to Automotive  & Ratio &  0.08 & 5\\
  \hline
  Pipe and Tubing to Machinery & Ratio & 0.04  & 5 \\
  \hline
  Pipe and Tubing to Energy & Ratio & 0.55  & 5\\
  \hline
  Pipe and Tubing to Export  & Ratio & 0.065  & 5\\
  \hline
  Plates to Construction & Ratio &  0.0408 & 5\\
  \hline
  Plates to Automotive  & Ratio &  0.01 & 5\\
  \hline
  Plates to Machinery & Ratio & 0.5187  & 5\\
  \hline
  Plates to Energy & Ratio & 0.067  & 5\\
  \hline
  Plates to Export  & Ratio &  0.231  & 5\\
  \hline
  Bars to Construction &  Ratio & 0.152 &  5\\
  \hline
  Bars to Automotive  & Ratio & 0.311  & 5\\
  \hline
  Bars to Machinery & Ratio &  0.238 & 5\\
  \hline
  Bars to Energy & Ratio & 0.046  & 5\\
  \hline
  Bars to Export  & Ratio & 0.131  & 5\\
  \hline
  Reinforcing Bars to Construction & Ratio & 0.925  & 5 \\
  \hline
  Reinforcing Bars to Export       & Ratio &  0.039 & 5 \\
  \hline
  Tin Mill Products to Automotive  & Ratio & 0.006  & 5 \\
  \hline
  Tin Mill Products to Steel Products & Ratio & 0.685  & 5\\
  \hline
  Tin Mill Products to Export       & Ratio &  0.067 & 5\\
  \hline
  Wire and Wire Rods to Construction & Ratio & 0.388  & 5\\
  \hline
  Wire and Wire Rods to Automotive  & Ratio & 0.285 & 5\\
  \hline
  Wire and Wire Rods to Machinery & Ratio & 0.1  &  5\\
  \hline
  Wire and Wire Rods to Energy & Ratio &  0.049 & 5\\
  \hline
  Wire and Wire Rods to Export  & Ratio & 0.094  & 5\\
  \hline
  Rail and Rail Accessories to Construction & Ratio &  0.779 & 5\\
  \hline
  Rail and Rail Accessories to Machinery & Ratio & 0.047  & 5 \\
  \hline
  Rail and Rail Accessories to Export  & Ratio & 0.141 & 5 \\
  \hline
  Light Section to Construction & Ratio &  0.86 & 5 \\
  \hline
  Light Section to Automotive & Ratio & 0.026  & 5\\
  \hline
  Light Section to Export  & Ratio & 0.057  & 5\\
 \hline
  Heavy Section to Construction & Ratio &  0.877 & 5\\
  \hline
  Heavy Section to Export & Ratio & 0.092  & 5\\
  \hline
  Steel Product Casting to Construction & Ratio & 0.259  & 5\\
  \hline
  Steel Product Casting to Automotive  & Ratio & 0.385  & 5\\
 \hline
  Steel Product Casting to Machinery & Ratio & 0.259  & 5\\
  \hline
  Steel Product Casting to Export &  Ratio & 0.111  & 5\\
  \hline
  Iron Product Casting to Construction & Ratio & 0.311  & 5\\
  \hline
  Iron Product Casting to Automotive  & Ratio & 0.552  & 5 \\
 \hline
  Iron Product Casting to Machinery & Ratio & 0.066  & 5 \\
  \hline
  Iron Product Casting to Export & Ratio & 0.07  & 5\\
  \hline
 \end{longtable}
\label{t:SI_2012 data}
\end{center}

\begin{table}[hb]
\begin{tabular}{cc}
 \hline
 \multicolumn{2}{l} {\textbf{Reference in Table 3}} \\
 \hline\hline
 1  & \makecell[l]{USGS. 2012. Iron Ore. Minerals Yearbook.  https://www.usgs.gov/centers/ \\ national-minerals-information-center/iron-ore-statistics-and-information}\\
 \hline
 2 & \makecell[l]{USGS. 2012. Iron and Steel Scrap. Minerals Yearbook. https://www.usgs.gov/centers/\\national-minerals-information-center/iron-and-steel-scrap-statistics-and-information} \\
  \hline
 3 & \makecell[l]{USGS. 2012. Iron and Steel. Minerals Yearbook. https://www.usgs.gov/centers/\\national-minerals-information-center/iron-and-steel-statistics-and-information} \\
  \hline
 4 & \makecell[l]{WorldSteel. 2017. Steel Statistical Yearbook 2017. https://worldsteel.org/steel-by-topic/\\statistics/steel-statistical-yearbook/} \\
  \hline
 5 & \makecell[l]{Yongxian Zhu, Kyle Syndergaard, and Daniel R. Cooper. Environmental Science \\ \& Technology  2019 53 (19) 11260-11268. DOI: 10.1021/acs.est.9b01016}\\
  \hline
 \end{tabular}
\end{table}

\subsection{Sankey diagram for all 16 candidate network structures}
\label{SI:sankey}

Figures below show the prior and posterior mass flows as Sankey diagrams of all 16 candidate network structures for the 2012 U.S. steel flow.

\begin{figure}
    \centering
    \includegraphics[width=\textwidth,height=\textheight,keepaspectratio]{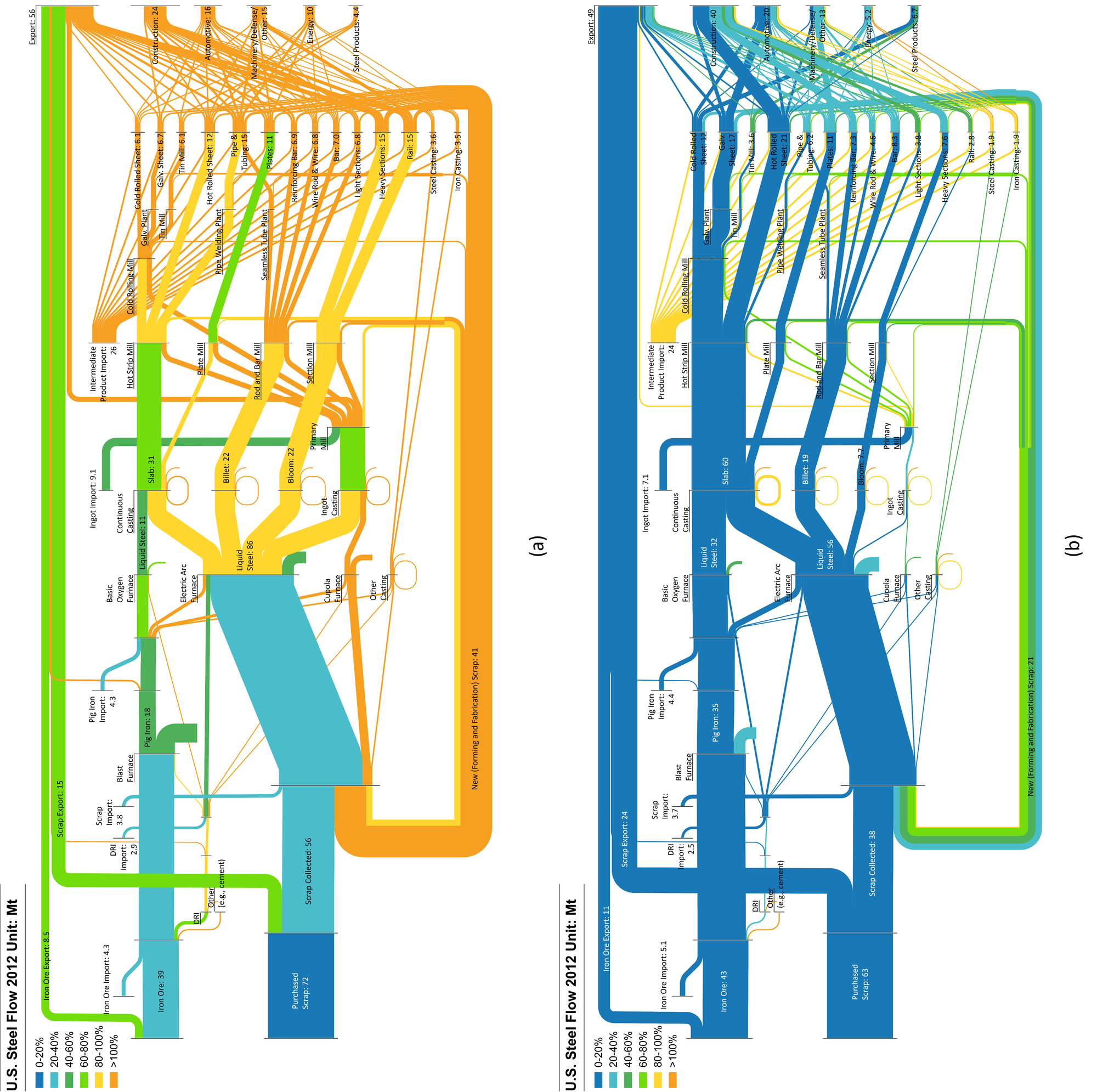}
    \caption{Bayesian (a) prior- and (b) posterior-predictive mass flows of network structure 0000 for the U.S. steel flow in 2012.
    All numbers on the flows refer to the mean of the mass flow in units of million metric tons (Mt). The uncertainty percentages refer to the flow standard deviation as a percentage of the mean of the mass flow. All mass flows refer to steel except for the iron ore flows that include the non-iron mass (e.g., oxygen and gangue).}
    \label{f:SI_0000}
\end{figure}

\begin{figure}
    \centering
    \includegraphics[width=\textwidth,height=\textheight,keepaspectratio]{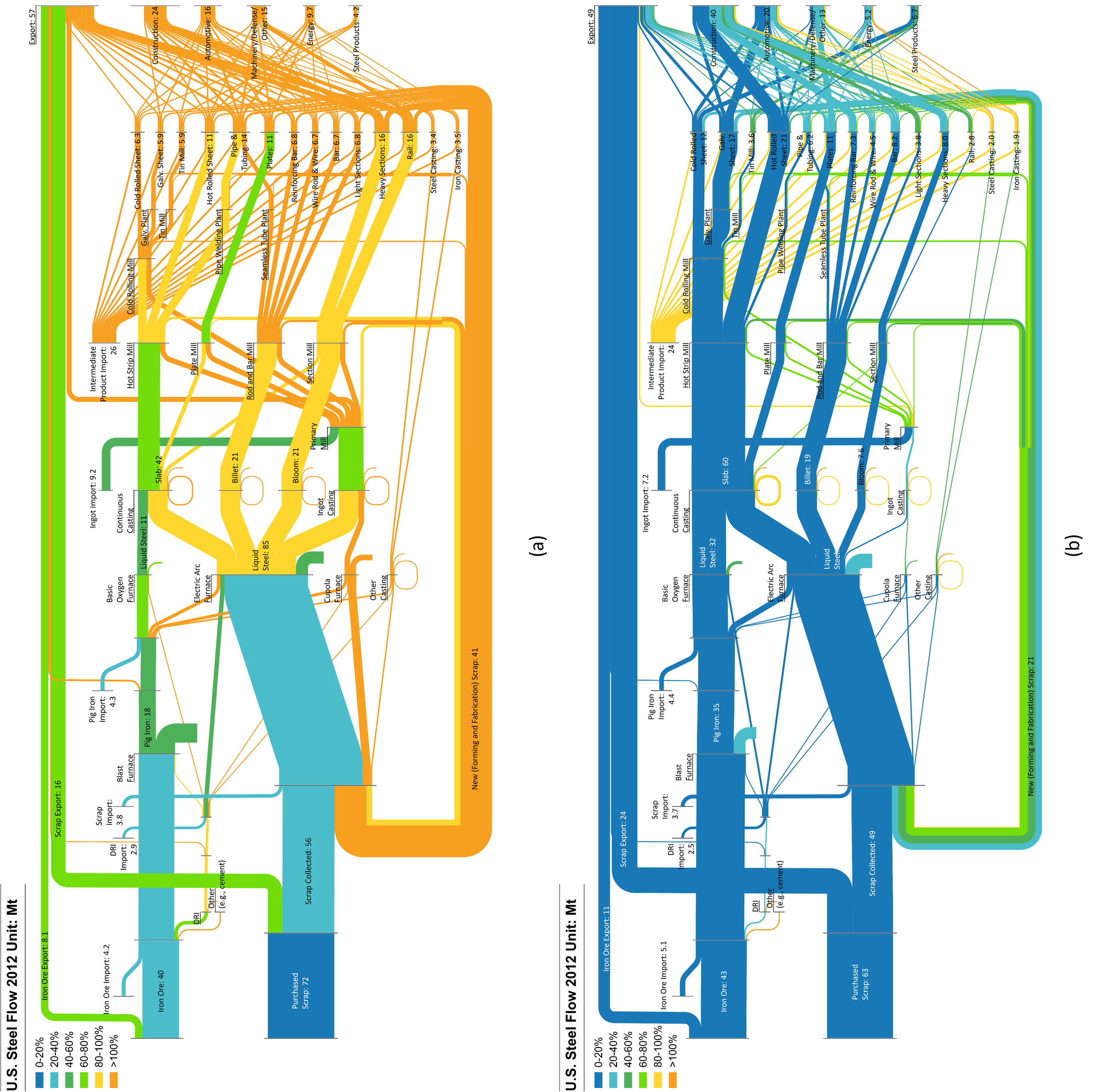}
    \caption{Bayesian (a) prior- and (b) posterior-predictive mass flows of network structure 0001 for the U.S. steel flow in 2012.
    All numbers on the flows refer to the mean of the mass flow in units of million metric tons (Mt). The uncertainty percentages refer to the flow standard deviation as a percentage of the mean of the mass flow. All mass flows refer to steel except for the iron ore flows that include the non-iron mass (e.g., oxygen and gangue).}
    \label{f:SI_0001}
\end{figure}

\begin{figure}
    \centering
    \includegraphics[width=\textwidth,height=\textheight,keepaspectratio]{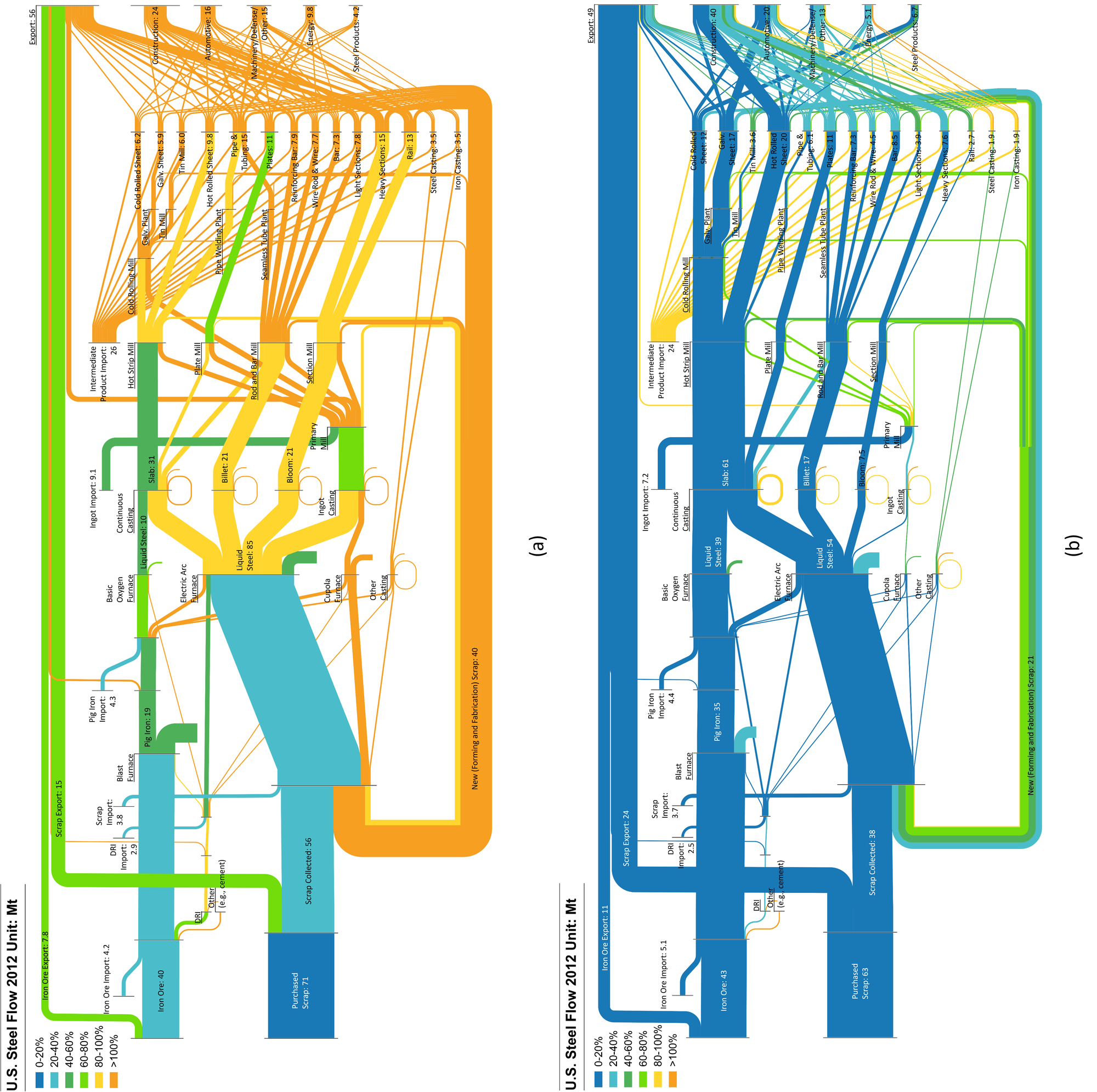}
    \caption{Bayesian (a) prior- and (b) posterior-predictive mass flows of network structure 0010 for the U.S. steel flow in 2012.
    All numbers on the flows refer to the mean of the mass flow in units of million metric tons (Mt). The uncertainty percentages refer to the flow standard deviation as a percentage of the mean of the mass flow. All mass flows refer to steel except for the iron ore flows that include the non-iron mass (e.g., oxygen and gangue).}
    \label{f:SI_0010}
\end{figure}

\begin{figure}
    \centering
    \includegraphics[width=\textwidth,height=\textheight,keepaspectratio]{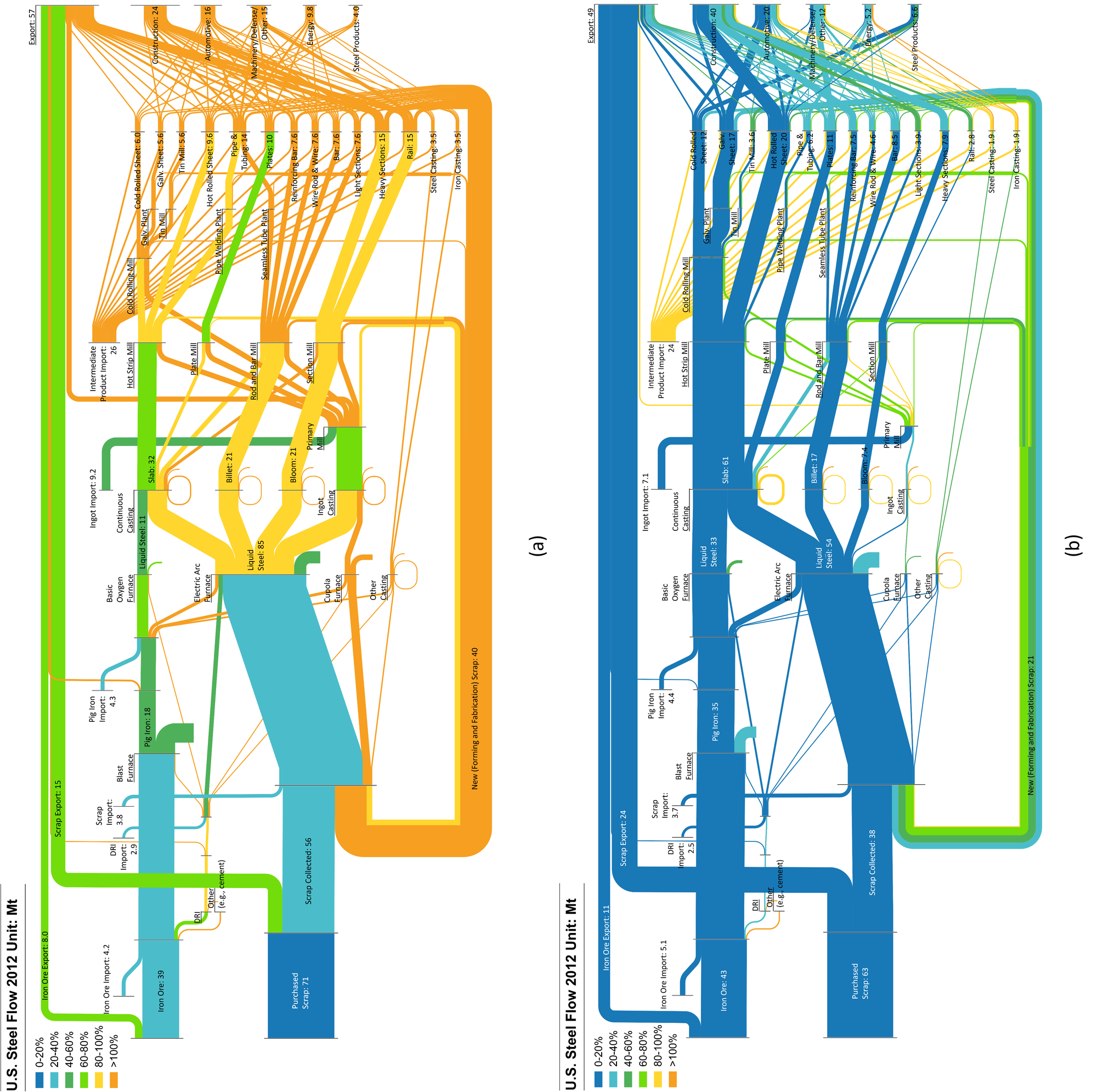}
    \caption{Bayesian (a) prior- and (b) posterior-predictive mass flows of network structure 0011 for the U.S. steel flow in 2012.
    All numbers on the flows refer to the mean of the mass flow in units of million metric tons (Mt). The uncertainty percentages refer to the flow standard deviation as a percentage of the mean of the mass flow. All mass flows refer to steel except for the iron ore flows that include the non-iron mass (e.g., oxygen and gangue).}
    \label{f:SI_0011}
\end{figure}

\begin{figure}
    \centering
    \includegraphics[width=\textwidth,height=\textheight,keepaspectratio]{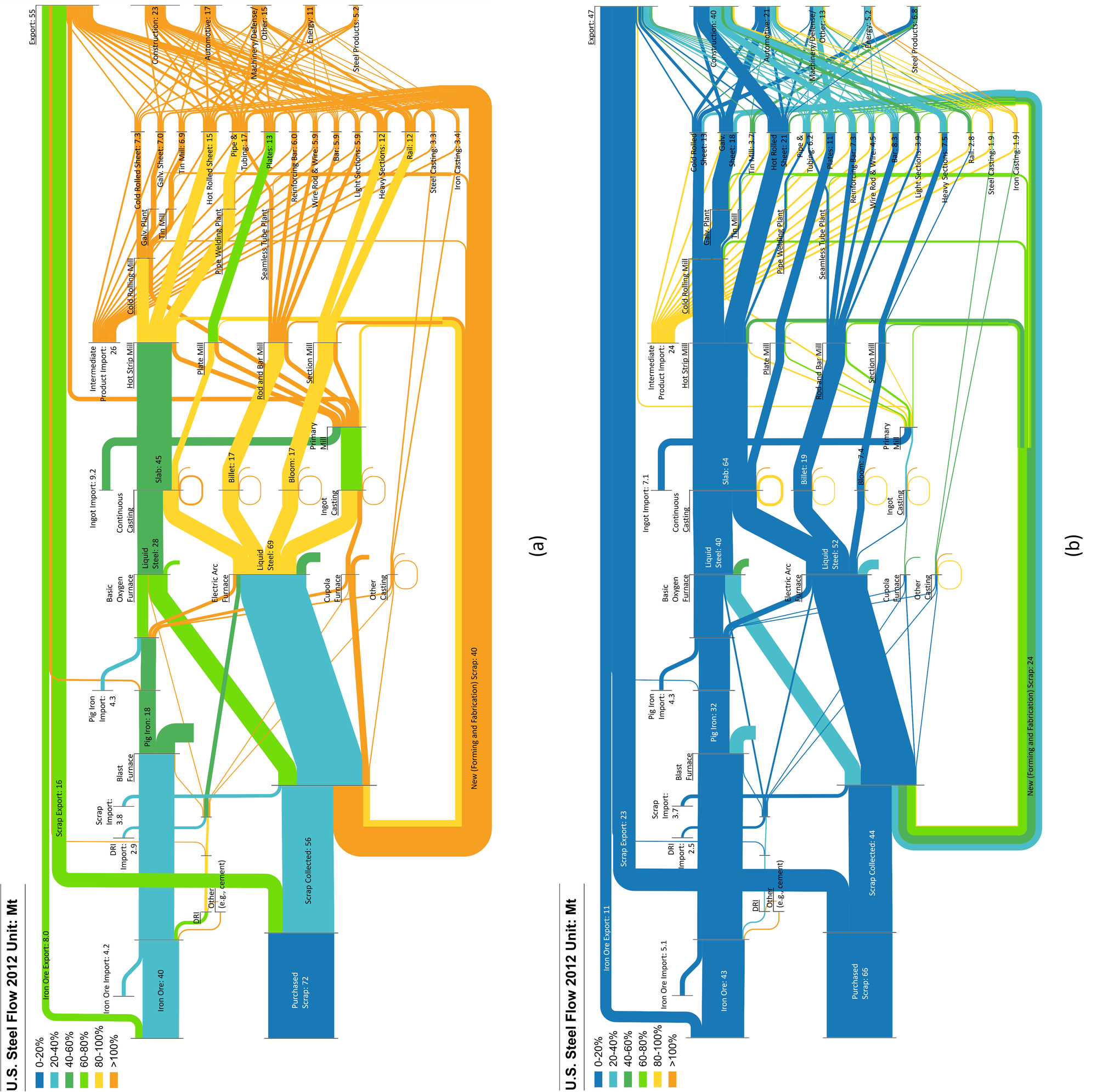}
    \caption{Bayesian (a) prior- and (b) posterior-predictive mass flows of network structure 0100 for the U.S. steel flow in 2012.
    All numbers on the flows refer to the mean of the mass flow in units of million metric tons (Mt). The uncertainty percentages refer to the flow standard deviation as a percentage of the mean of the mass flow. All mass flows refer to steel except for the iron ore flows that include the non-iron mass (e.g., oxygen and gangue).}
    \label{f:SI_0100}
\end{figure}

\begin{figure}
    \centering
    \includegraphics[width=\textwidth,height=\textheight,keepaspectratio]{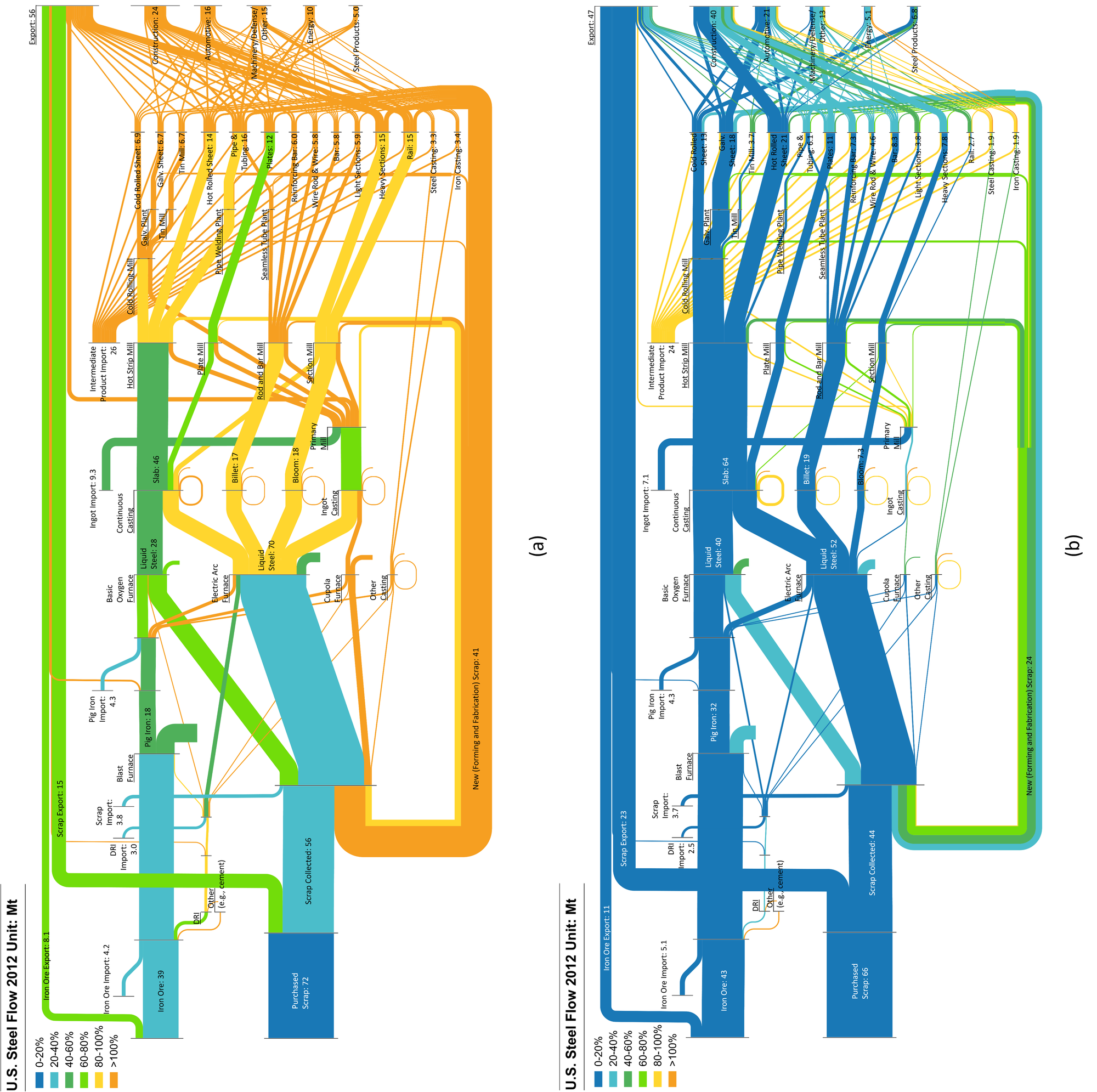}
    \caption{Bayesian (a) prior- and (b) posterior-predictive mass flows of network structure 0101 for the U.S. steel flow in 2012.
    All numbers on the flows refer to the mean of the mass flow in units of million metric tons (Mt). The uncertainty percentages refer to the flow standard deviation as a percentage of the mean of the mass flow. All mass flows refer to steel except for the iron ore flows that include the non-iron mass (e.g., oxygen and gangue).}
    \label{f:SI_0101}
\end{figure}

\begin{figure}
    \centering
    \includegraphics[width=\textwidth,height=\textheight,keepaspectratio]{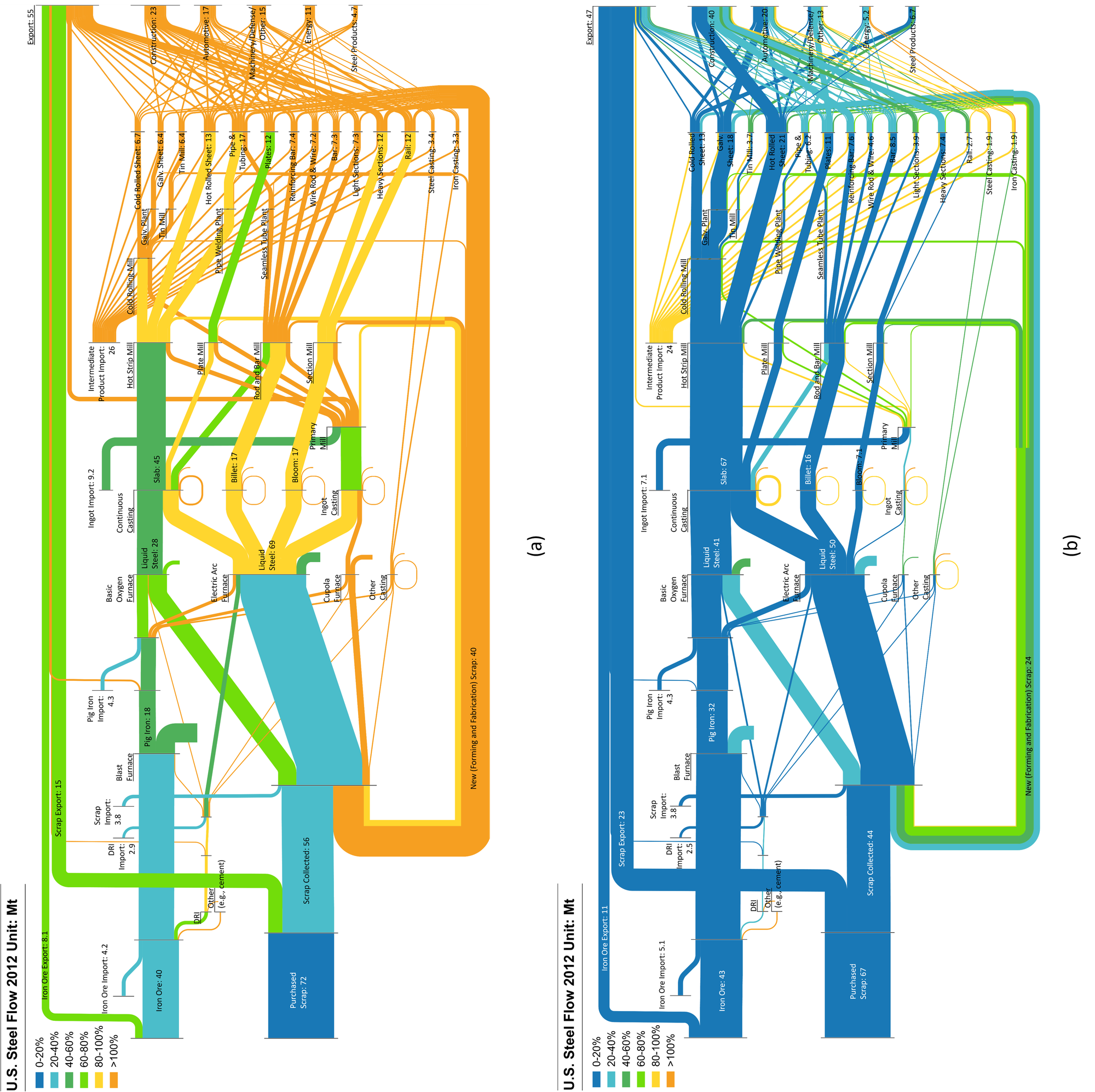}
    \caption{Bayesian (a) prior- and (b) posterior-predictive mass flows of network structure 0110 for the U.S. steel flow in 2012.
    All numbers on the flows refer to the mean of the mass flow in units of million metric tons (Mt). The uncertainty percentages refer to the flow standard deviation as a percentage of the mean of the mass flow. All mass flows refer to steel except for the iron ore flows that include the non-iron mass (e.g., oxygen and gangue).}
    \label{f:SI_0110}
\end{figure}

\begin{figure}
    \centering
    \includegraphics[width=\textwidth,height=\textheight,keepaspectratio]{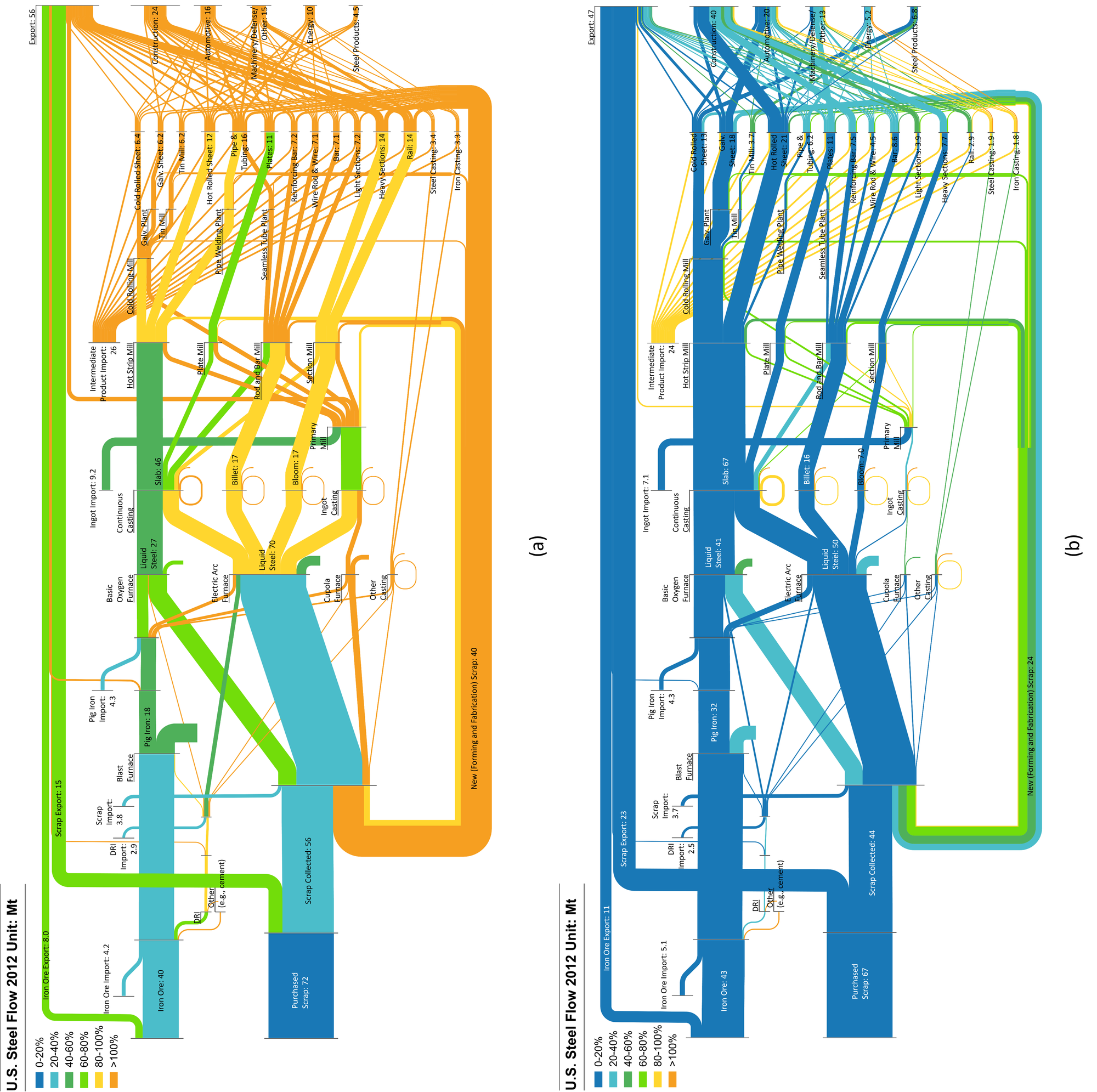}
    \caption{Bayesian (a) prior- and (b) posterior-predictive mass flows of network structure 0111 for the U.S. steel flow in 2012.
    All numbers on the flows refer to the mean of the mass flow in units of million metric tons (Mt). The uncertainty percentages refer to the flow standard deviation as a percentage of the mean of the mass flow. All mass flows refer to steel except for the iron ore flows that include the non-iron mass (e.g., oxygen and gangue).}
    \label{f:SI_0111}
\end{figure}

\begin{figure}
    \centering
    \includegraphics[width=\textwidth,height=\textheight,keepaspectratio]{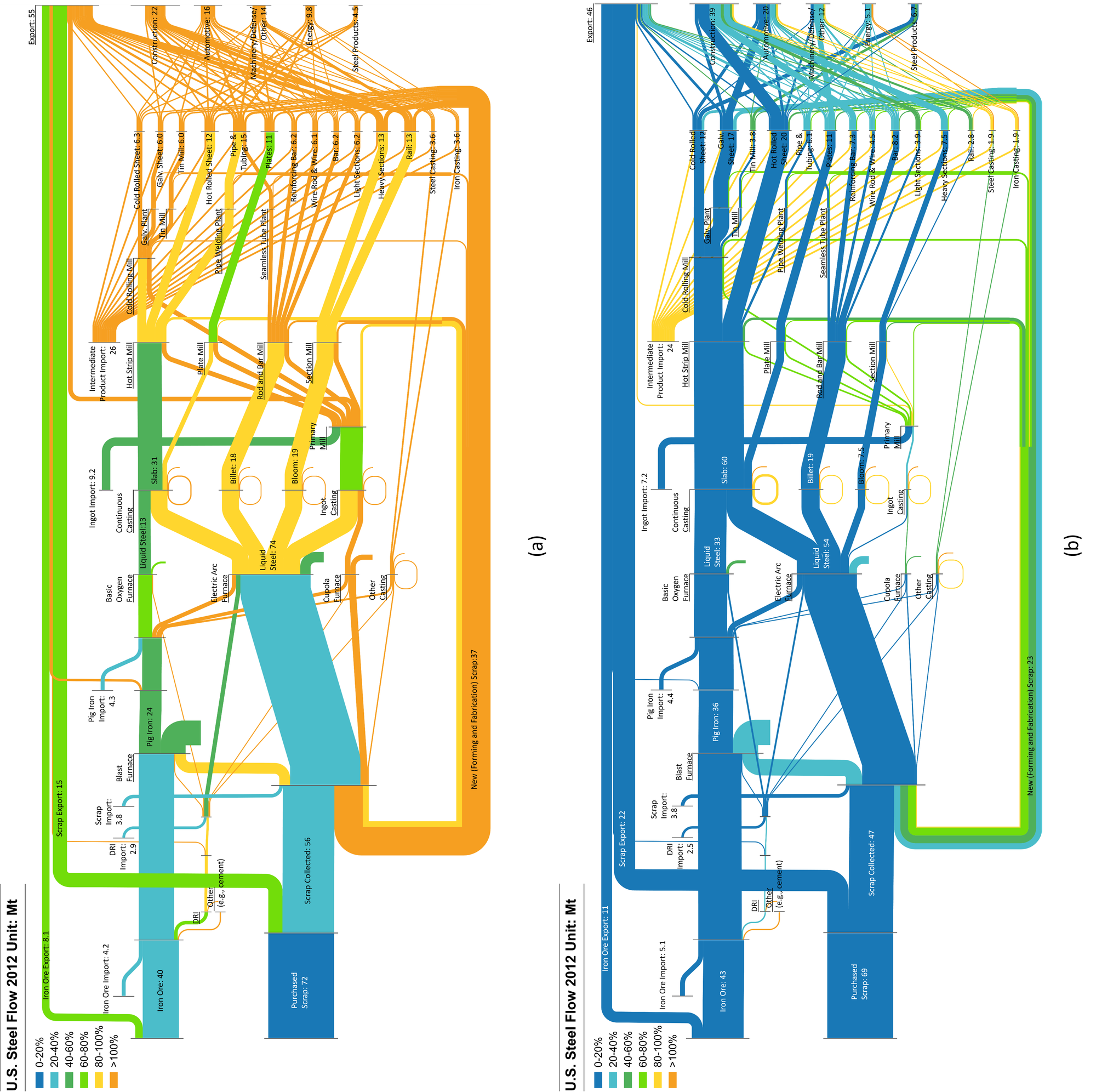}
    \caption{Bayesian (a) prior- and (b) posterior-predictive mass flows of network structure 1000 for the U.S. steel flow in 2012.
    All numbers on the flows refer to the mean of the mass flow in units of million metric tons (Mt). The uncertainty percentages refer to the flow standard deviation as a percentage of the mean of the mass flow. All mass flows refer to steel except for the iron ore flows that include the non-iron mass (e.g., oxygen and gangue).}
    \label{f:SI_1000}
\end{figure}

\begin{figure}
    \centering
    \includegraphics[width=\textwidth,height=\textheight,keepaspectratio]{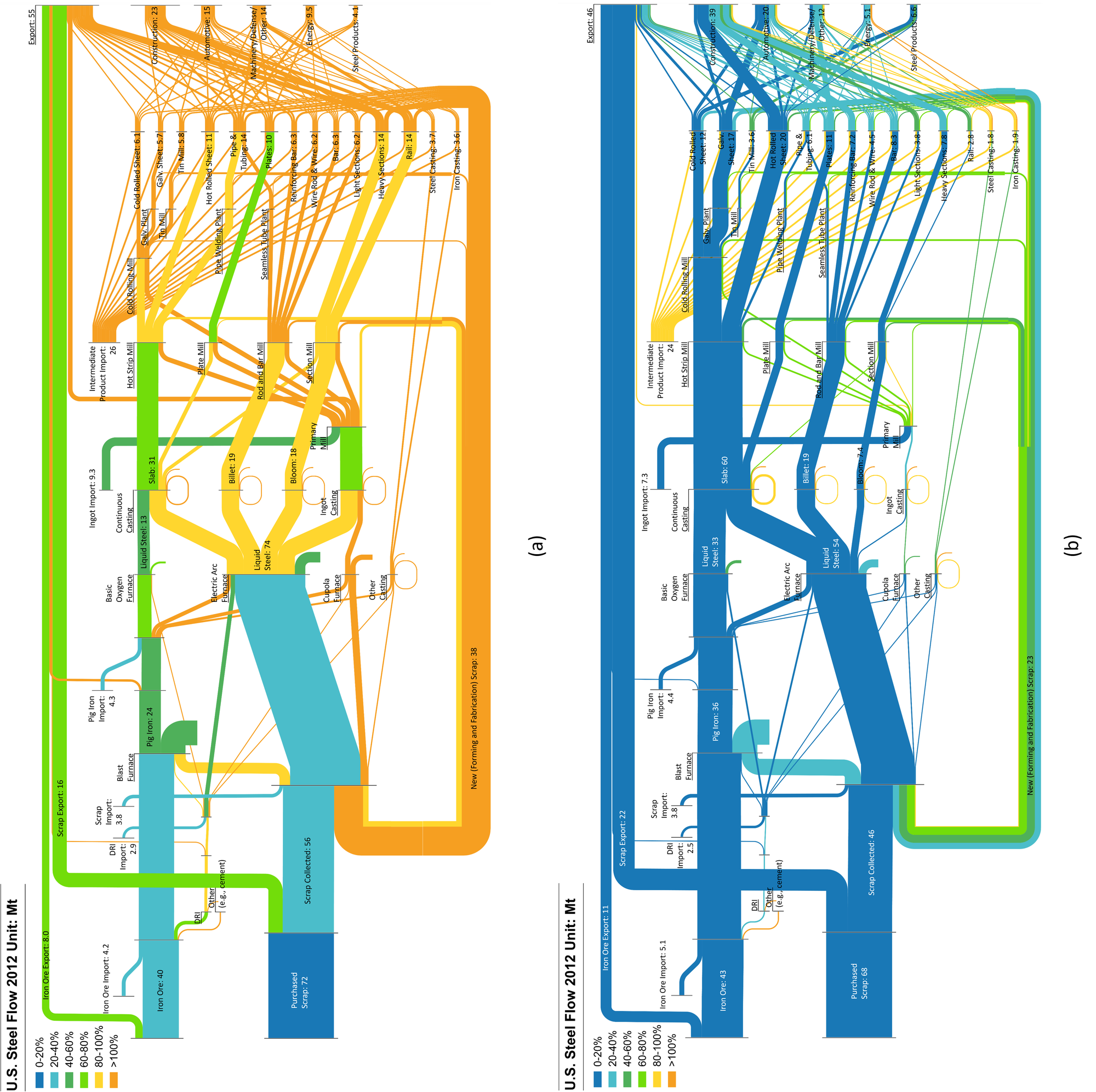}
    \caption{Bayesian (a) prior- and (b) posterior-predictive mass flows of network structure 1001 for the U.S. steel flow in 2012.
    All numbers on the flows refer to the mean of the mass flow in units of million metric tons (Mt). The uncertainty percentages refer to the flow standard deviation as a percentage of the mean of the mass flow. All mass flows refer to steel except for the iron ore flows that include the non-iron mass (e.g., oxygen and gangue).}
    \label{f:SI_1001}
\end{figure}

\begin{figure}
    \centering
    \includegraphics[width=\textwidth,height=\textheight,keepaspectratio]{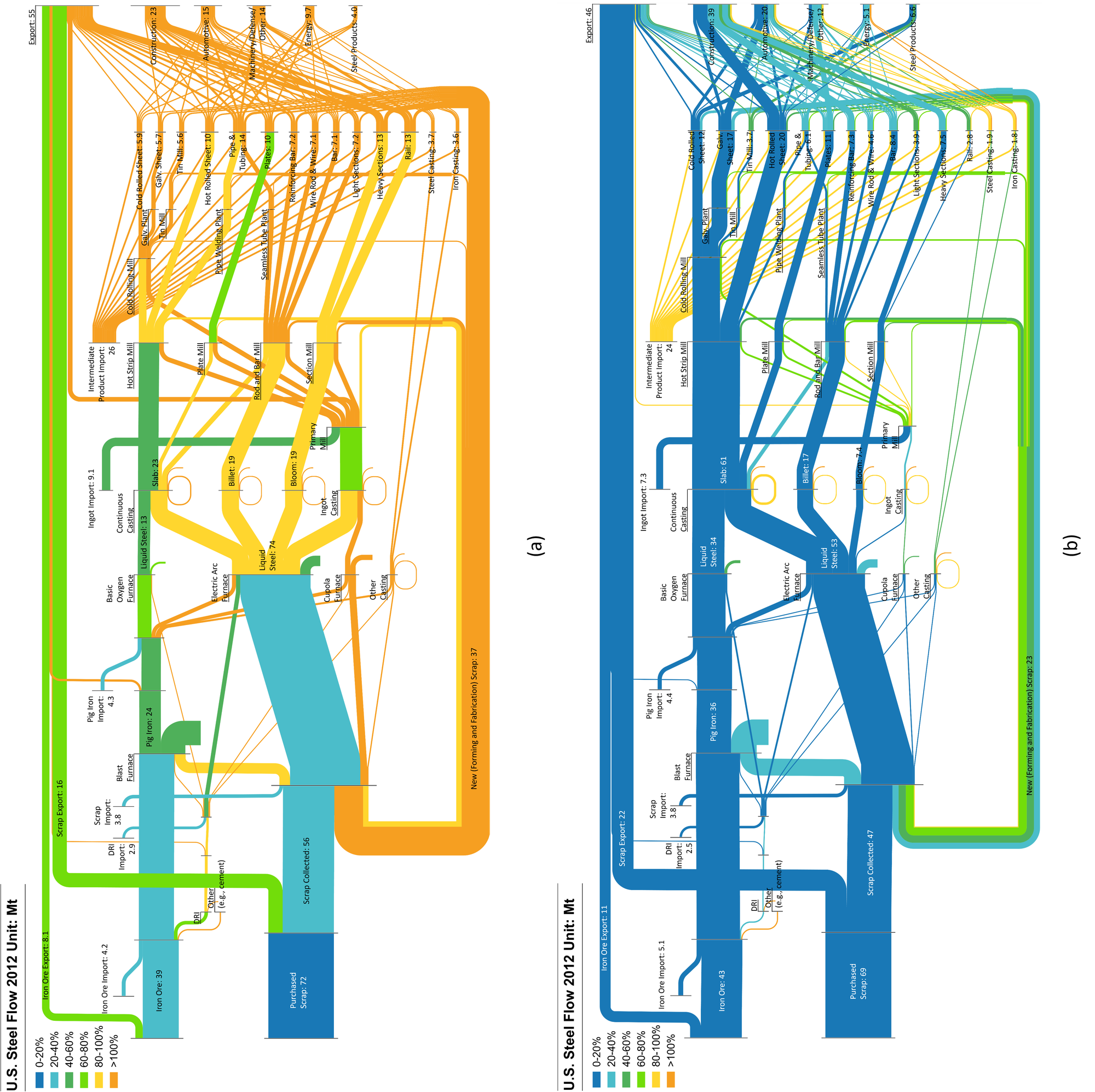}
    \caption{Bayesian (a) prior- and (b) posterior-predictive mass flows of network structure 1010 for the U.S. steel flow in 2012.
    All numbers on the flows refer to the mean of the mass flow in units of million metric tons (Mt). The uncertainty percentages refer to the flow standard deviation as a percentage of the mean of the mass flow. All mass flows refer to steel except for the iron ore flows that include the non-iron mass (e.g., oxygen and gangue).}
    \label{f:SI_1010}
\end{figure}

\begin{figure}
    \centering
    \includegraphics[width=\textwidth,height=\textheight,keepaspectratio]{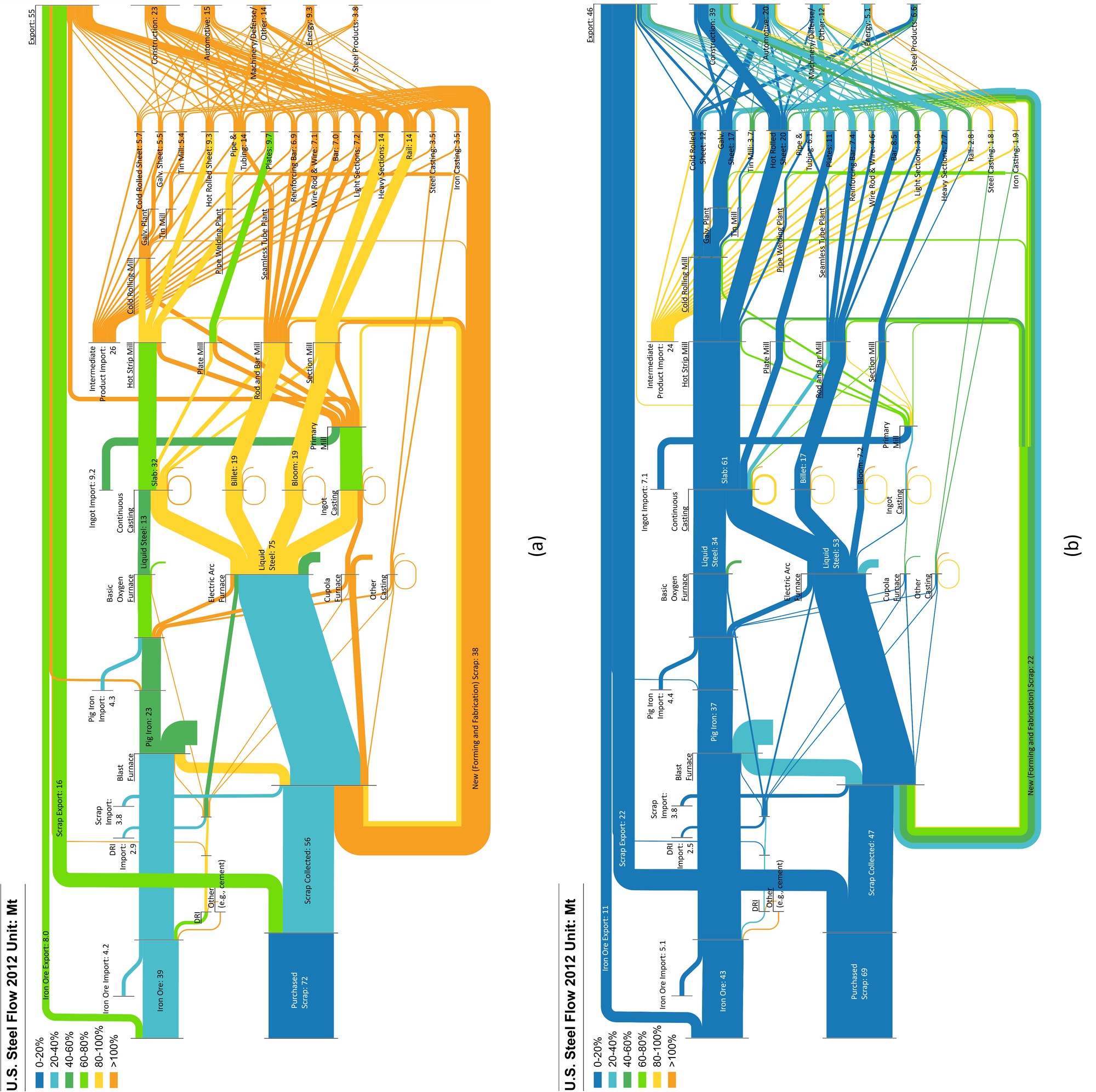}
    \caption{Bayesian (a) prior- and (b) posterior-predictive mass flows of network structure 1011 for the U.S. steel flow in 2012.
    All numbers on the flows refer to the mean of the mass flow in units of million metric tons (Mt). The uncertainty percentages refer to the flow standard deviation as a percentage of the mean of the mass flow. All mass flows refer to steel except for the iron ore flows that include the non-iron mass (e.g., oxygen and gangue).}
    \label{f:SI_1011}
\end{figure}

\begin{figure}
    \centering
    \includegraphics[width=\textwidth,height=\textheight,keepaspectratio]{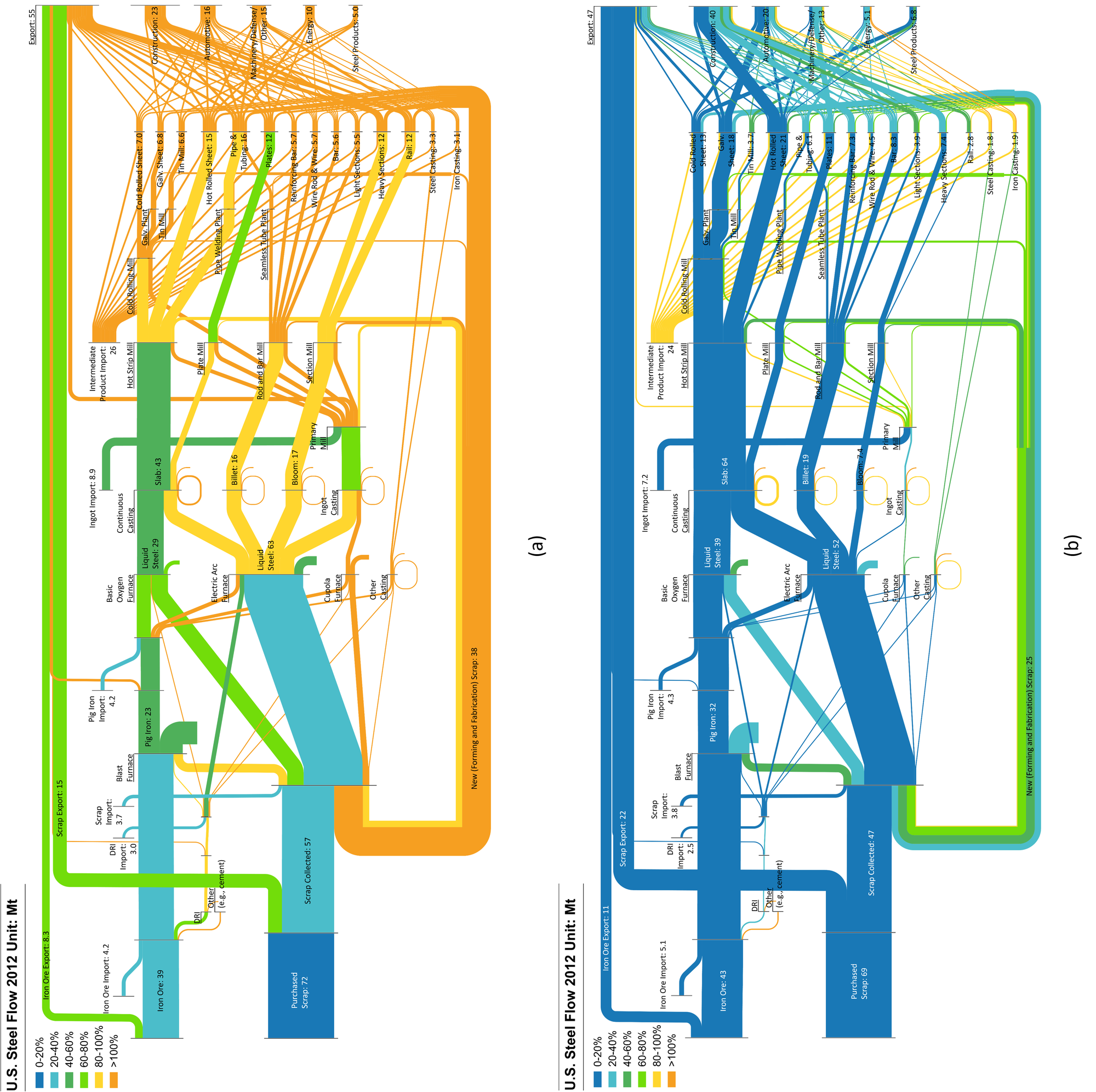}
    \caption{Bayesian (a) prior- and (b) posterior-predictive mass flows of network structure 1100 for the U.S. steel flow in 2012.
    All numbers on the flows refer to the mean of the mass flow in units of million metric tons (Mt). The uncertainty percentages refer to the flow standard deviation as a percentage of the mean of the mass flow. All mass flows refer to steel except for the iron ore flows that include the non-iron mass (e.g., oxygen and gangue).}
    \label{f:SI_1100}
\end{figure}

\begin{figure}
    \centering
    \includegraphics[width=\textwidth,height=\textheight,keepaspectratio]{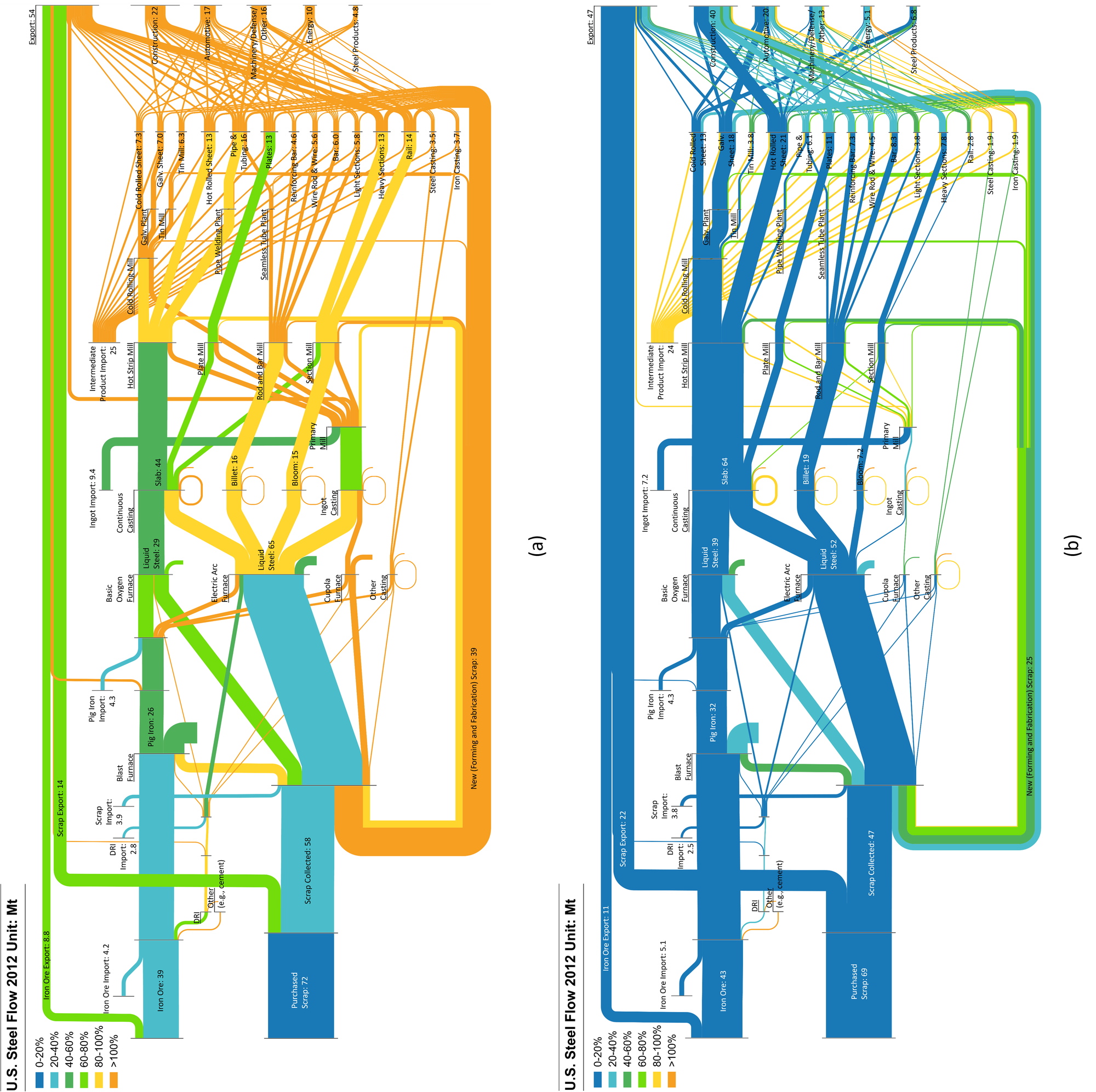}
    \caption{Bayesian (a) prior- and (b) posterior-predictive mass flows of network structure 1101 for the U.S. steel flow in 2012.
    All numbers on the flows refer to the mean of the mass flow in units of million metric tons (Mt). The uncertainty percentages refer to the flow standard deviation as a percentage of the mean of the mass flow. All mass flows refer to steel except for the iron ore flows that include the non-iron mass (e.g., oxygen and gangue).}
    \label{f:SI_1101}
\end{figure}

\begin{figure}
    \centering
    \includegraphics[width=\textwidth,height=\textheight,keepaspectratio]{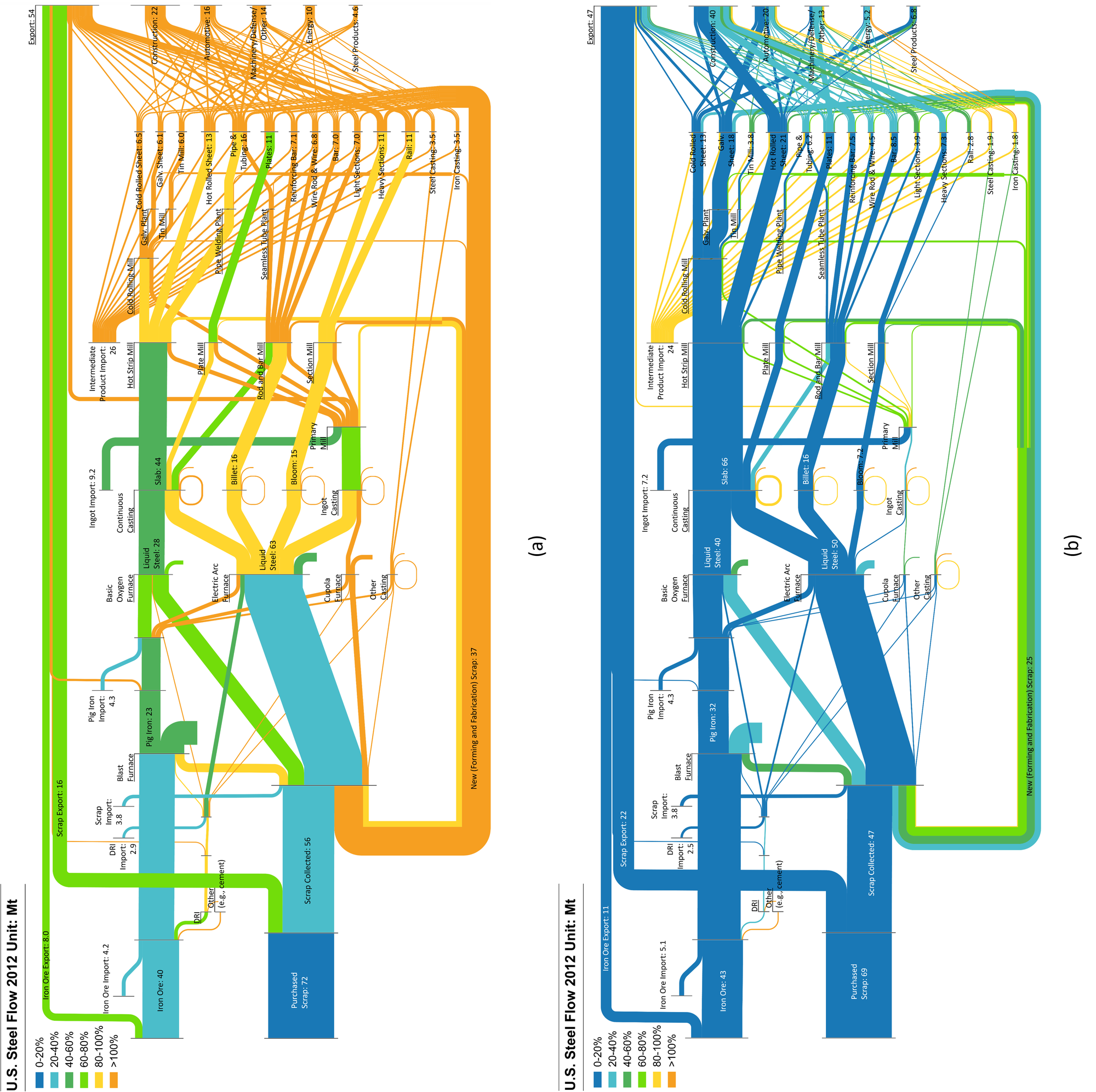}
    \caption{Bayesian (a) prior- and (b) posterior-predictive mass flows of network structure 1110 for the U.S. steel flow in 2012.
    All numbers on the flows refer to the mean of the mass flow in units of million metric tons (Mt). The uncertainty percentages refer to the flow standard deviation as a percentage of the mean of the mass flow. All mass flows refer to steel except for the iron ore flows that include the non-iron mass (e.g., oxygen and gangue).}
    \label{f:SI_1110}
\end{figure}

\begin{figure}
    \centering
    \includegraphics[width=\textwidth,height=\textheight,keepaspectratio]{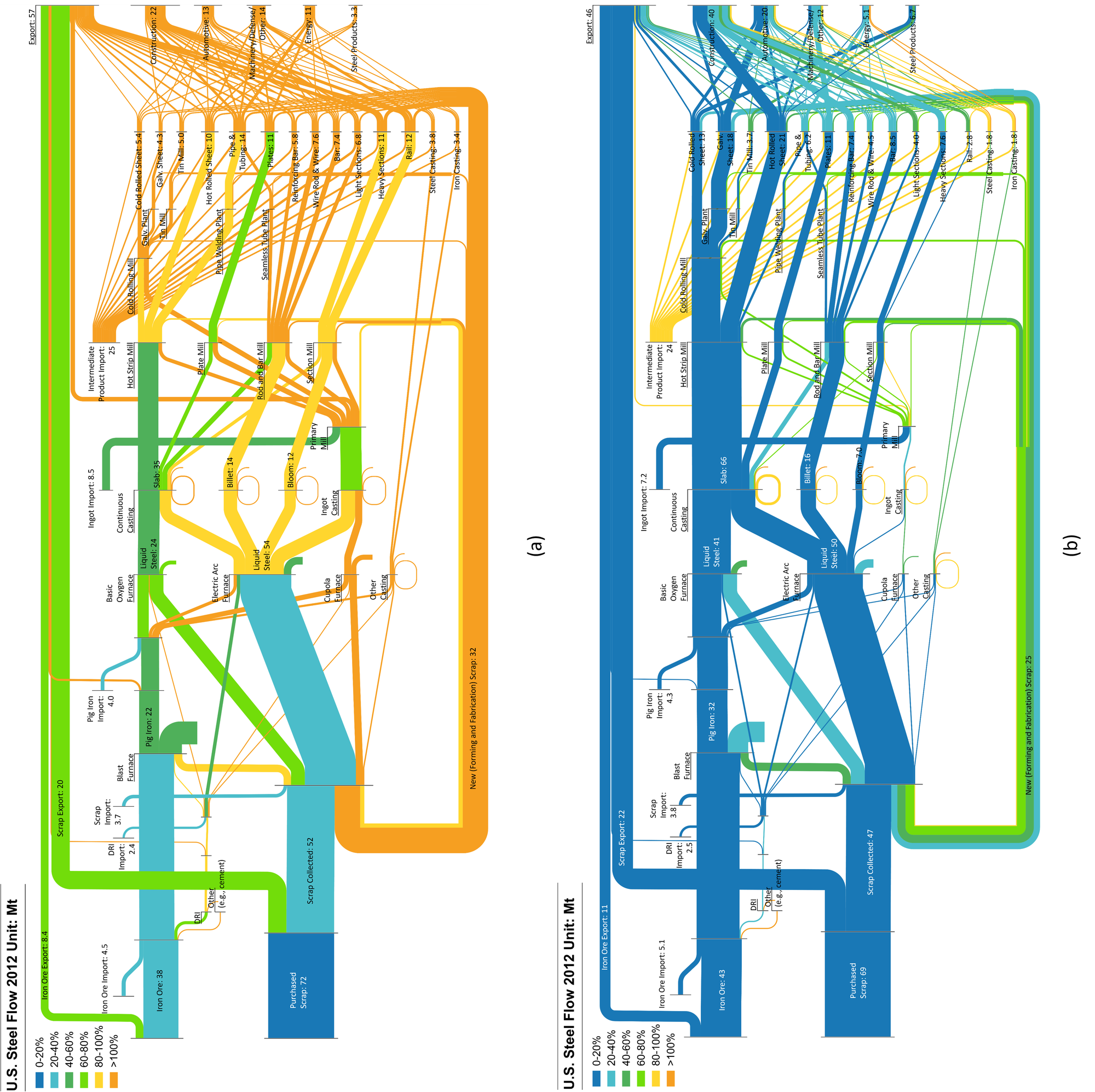}
    \caption{Bayesian (a) prior- and (b) posterior-predictive mass flows of network structure 1111 for the U.S. steel flow in 2012.
    All numbers on the flows refer to the mean of the mass flow in units of million metric tons (Mt). The uncertainty percentages refer to the flow standard deviation as a percentage of the mean of the mass flow. All mass flows refer to steel except for the iron ore flows that include the non-iron mass (e.g., oxygen and gangue).}
    \label{f:SI_1111}
\end{figure}

\subsection{Nodal emission intensities} 
\label{SI:emission}

\Cref{t:SI_node_emission} presents the estimated nodal emission intensities (kg.CO\textsubscript{2eq.}/kg.mat in) for the U.S. steel sector. These emission intensities were provided by Dr. Mohammad Heidari \cite{Heidari24}. We are grateful to him for this contribution. This analysis focuses on domestic emissions; i.e., emissions released within the U.S. Therefore, any import product (e.g., Import DRI) is attributed a nodal emission intensity factor of 0 kg.CO\textsubscript{2eq.}/kg.mat in. Some of the nodes in \cref{t:SI_node_emission} are compiler nodes. These nodes are for visualization and calculation purposes. They do not represent actual processes and therefore have an emission intensity of 0.

\begin{center}
\begin{longtable}{m{17em} m{4cm} m{5cm} }
    \caption{Node emission intensity as per unit of material into the process.}\\
     \hline
     Node name & Emission intensity [kg.CO\textsubscript{2}/kg.mat in] & Note\\
     \hline\hline
     Iron ore production & 0.11 & Domestic production of iron ore\\
     \hline
     Iron ore consumption & 0 & Compiler node aggregating imported iron ore and domestic iron ore not exported\\
     \hline
     Import iron ore & 0 & Focus of the analysis is domestic emission: imports assigned 0 emission intensities\\
     \hline
     DRI production & 0.67 & \\
     \hline
     DRI & 0 & Compiler node describing DRI produced domestically\\
     \hline
     Import DRI & 0 & Focus of the analysis is domestic emission: imports assigned 0 emission intensities\\
     \hline
     DRI consumption & 0 & Compiler node aggregating imported DRI and domestically produced DRI not exported\\
     \hline
     Blast furnace & 1.50\\
     \hline
     Import pig iron & 0 & Focus of the analysis is domestic emission: imports assigned 0 emission intensities\\
     \hline
     Pig iron & 0 & Compiler node describing pig iron produced domestically from blast furnace\\
     \hline
     Pig iron consumption & 0 & Compiler node aggregating imported pig iron and domestically pig iron not exported\\
     \hline
     Purchased scrap & 0.04 & Post-consumer scrap collected domestically\\
     \hline
     Scrap collected & 0 & Compiler node aggregating all post-consumer scrap collected domestically\\
     \hline
     Import scrap & 0 & Focus of the analysis is domestic emission: imports assigned 0 emission intensities\\
     \hline
     Scrap consumption & 0 & Node aggregating post-industrial process scraps and domestically collected post consumer scrap not exported\\
     \hline
     Basic oxygen furnace & 0.13\\
     \hline
     Electric Arc furnace & 0.29\\
     \hline
     EAF\_yield & 0 & Compiler node aggregating all products from electric arc furnace\\
     \hline
     Cupola furnace & 0.13 & Zhu \textit{et al.} \cite{Zhu23}\\
     \hline
     Other casting & 0.13 & Emission intensity modeled the same as continuous casting\\
     \hline
     OC\_yield & 0 & Compiler node aggregating all products from other casting process\\
     \hline
     OC\_loss & 0 & Compiler node aggregating run-around prep and loss for other casting process\\
     \hline
     Continuous casting - slabs & 0.13 & Continuous casting process producing slabs\\
     \hline
     CC\_yield & 0 & Compiler node aggregating all products from continuous cast slabs\\
     \hline
     CC\_loss & 0 & Compiler node aggregating run-around prep and loss for continuous cast slab\\
     \hline
     Continuous casting - billets & 0.13 & Continuous casting process producing billets\\
     \hline
     BT\_yield & 0 & Compiler node aggregating all products from continuous cast billets\\
     \hline
     BT\_loss & 0 & Compiler node aggregating run-around prep and loss for continuous cast billets\\
     \hline 
     Continuous casting - blooms & 0.13 & Continuous casting process producing blooms\\
     \hline
     BM\_yield & 0 & Compiler node aggregating all products from continuous cast blooms\\
     \hline
     BM\_loss & 0 & Compiler node aggregating run-around prep and loss for continuous cast blooms\\
     \hline
     Ingot casting & 0.13 & Emission intensity modeled the same as continuous casting\\
     \hline
     IC\_yield & 0 & Compiler node aggregating all products from ingot casting process\\
     \hline
     IC\_loss & 0 & Compiler node aggregating run-around prep and loss for ingot casting process\\
     \hline
     Ingot import & 0 & Focus of the analysis is domestic emission: imports assigned 0 emission intensities\\
     \hline
     Primary mill & 0.13 & Emission intensity modeled the same as continuous casting\\
     \hline
     PM\_Yield & 0 & Compiler node aggregating all products from primary mill\\
     \hline
     Hot strip mill & 0.18 & An emission intensity of 0.25 kg.co\textsubscript{2eq.}/kg.mat in \cite{Heidari24} modeled for the combined hot and cold rolling process; Milford \textit{et al.} \cite{Milford11} reports a 72--28 emission intensity split between hot and cold rolling process\\
     \hline
     HSM\_Yield & 0 & Compiler node aggregating all products from hot strip mill\\
     \hline
     Plate mill & 0.18 & Emission intensity modeled the same as hot rolling process\\
     \hline
     Rod and bar mill & 0.18 & Emission intensity modeled the same as hot rolling process\\
     \hline
     RBM\_Yield & 0 & Compiler node aggregating all products from rod and bar mill\\
     \hline
     Section mill & 0.18 & Emission intensity modeled the same as hot rolling process into i-beam, profiled rolling process\\
     \hline
     SM\_Yield & 0 & Compiler node aggregating all products from section mill\\
     \hline
     Cold rolling mill & 0.07 & An emission intensity of 0.25 kg.co\textsubscript{2eq.}/kg.mat in \cite{Heidari24} modeled for the combined hot and cold rolling process; Milford \textit{et al.} \cite{Milford11} reports a 72--28 emission intensity split between hot and cold rolling process\\
     \hline
     CRM\_Yield & 0 & Compiler node aggregating all products from cold rolling mill\\
     \hline
     Galvanizing plant & 0.19 & Galvanizing plant taking sheet rolls and coating with zinc\\
     \hline
     Tin mill & 0.08 & Emission intensity for tin mills adapted from galvanizing process, with the minimal energy to melt tin 40\% lower than that of zinc\\
     \hline
     Pipe and tubing & 0.18 & Emission intensity modeled the same as hot rolling process\\
     \hline
     Bars & 0 & Cutting process with negligible emission intensities\\
     \hline
     Cold rolled sheet & 0.32 & Manufacturing process of stamping and assembly of steel sheets\\
     \hline
     Galvanized sheet & 0.32 & Manufacturing process of stamping and assembly of steel sheets\\
     \hline
     Hot rolled sheet & 0.32 & Manufacturing process of stamping and assembly of steel sheets\\
     \hline
     Iron product casting & 0 & Machining process with negligible emission intensities at critical interfaces of intermediate products\\
     \hline
     Light section & 0 & Cutting process with negligible emission intensities\\
     \hline
     Pipe welding plant & 0.02 & GREET model\cite{ANL23} reports 32.6 kg.CO\textsubscript{2eq.} per passenger vehicle with an average vehicle weight of 1443 kg\\
     \hline
     Plates & 0 & Cutting process with negligible emission intensities\\
     \hline
     Seamless tube plant & 0.18 & Emission intensity modeled the same as hot rolling process\\
     \hline
     Reinforcing bars & 0 & Cutting process with negligible emission intensities\\
     \hline
     Rails and rail accessories & 0 & Cutting process with negligible emission intensities\\
     \hline
     Heavy section & 0 & Cutting process with negligible emission intensities\\
     \hline
     Tin mill products & 0.32 & Manufacturing process of stamping and assembly of steel sheets\\
     \hline
     Wire and wire rods & 0.01 & Forming process to manufacture wire, fasteners and tools\\
     \hline
     Steel product casting & 0 & Machining process with negligible emission intensities at critical interfaces of intermediate products\\
     \hline
     Intermediate product import & 0 & Focus of the analysis is domestic emission: imports assigned 0 emission intensities\\
     \hline
\label{t:SI_node_emission}
\end{longtable}
\end{center}


\bibliography{local}

\begin{thebibliography}{10}

\bibitem{Brunner16}
Paul~H Brunner and Helmut Rechberger.
\newblock {\em Handbook of material flow analysis: For environmental, resource,
  and waste engineers}.
\newblock CRC press, 2016.

\bibitem{Cullen22}
J.M. Cullen and D.R Cooper.
\newblock Material flows and uncertainty.
\newblock {\em Annual Review of Materials Research}, 2022.
\newblock \href
  {https://doi.org/https://doi.org/10.1146/annurev-matsci-070218-125903}
  {\path{doi:https://doi.org/10.1146/annurev-matsci-070218-125903}}.

\bibitem{Kopec16}
Grant~M. Kopec, Julian~M. Allwood, Jonathan~M. Cullen, and Daniel Ralph.
\newblock A general nonlinear least squares data reconciliation and estimation
  method for material flow analysis.
\newblock {\em Journal of Industrial Ecology}, 20(5):1038--1049, 2016.
\newblock URL:
  \url{https://onlinelibrary.wiley.com/doi/abs/10.1111/jiec.12344}, \href
  {http://arxiv.org/abs/https://onlinelibrary.wiley.com/doi/pdf/10.1111/jiec.12344}
  {\path{arXiv:https://onlinelibrary.wiley.com/doi/pdf/10.1111/jiec.12344}},
  \href {https://doi.org/https://doi.org/10.1111/jiec.12344}
  {\path{doi:https://doi.org/10.1111/jiec.12344}}.

\bibitem{Schwab18}
Oliver Schwab and Helmut Rechberger.
\newblock Information content, complexity, and uncertainty in material flow
  analysis.
\newblock {\em Journal of Industrial Ecology}, 22(2):263--274, 2018.
\newblock URL:
  \url{https://onlinelibrary.wiley.com/doi/abs/10.1111/jiec.12572?casa_token=rLjn8SaEB-gAAAAA:gsFsJYrbHeYjiGwRGspVJkOX0h19p7nopi0a2gbRrcQZyFWU0VdvQF4LoDVYaCSzwKXPvfZcfgOxgw},
  \href {https://doi.org/https://doi.org/10.1111/jiec.12572}
  {\path{doi:https://doi.org/10.1111/jiec.12572}}.

\bibitem{Graedel19}
Thomas Graedel.
\newblock Material flow analysis from origin to evolution.
\newblock {\em Environmental science \& technology}, 2019, 09 2019.
\newblock \href {https://doi.org/10.1021/acs.est.9b03413}
  {\path{doi:10.1021/acs.est.9b03413}}.

\bibitem{Cencic08}
O.~Cencic and H.~Rechberger.
\newblock Material flow analysis with software stan.
\newblock {\em Journal of Environmental Engineering and Management},
  18(1):3--7, 2008.

\bibitem{Cencic16}
Oliver Cencic.
\newblock Nonlinear data reconciliation in material flow analysis with software
  stan.
\newblock {\em Sustainable Environment Research}, 26(6):291--298, 2016.
\newblock \href {https://doi.org/https://doi.org/10.1016/j.serj.2016.06.002}
  {\path{doi:https://doi.org/10.1016/j.serj.2016.06.002}}.

\bibitem{UNcomtrade12}
{United Nations Comtrade Database}.
\newblock International trade statistics yearbook, 2012.
\newblock \url{https://comtrade.un.org/pb/downloads/2012/VolI2012.pdf}.

\bibitem{Zhu19}
Yongxian Zhu, Kyle Syndergaard, and Daniel~R. Cooper.
\newblock Mapping the annual flow of steel in the united states.
\newblock {\em Environmental Science \& Technology}, 53(19):11260--11268, 2019.
\newblock PMID: 31468962.
\newblock \href {http://arxiv.org/abs/https://doi.org/10.1021/acs.est.9b01016}
  {\path{arXiv:https://doi.org/10.1021/acs.est.9b01016}}, \href
  {https://doi.org/10.1021/acs.est.9b01016}
  {\path{doi:10.1021/acs.est.9b01016}}.

\bibitem{Gottschalk10}
Fadri Gottschalk, Roland~W. Scholz, and Bernd Nowack.
\newblock {Probabilistic material flow modeling for assessing the environmental
  exposure to compounds: Methodology and an application to engineered nano-TiO2
  particles}.
\newblock {\em Environmental Modelling \& Software}, 25(3):320--332, 2010.
\newblock \href {https://doi.org/10.1016/j.envsoft.2009.08.011}
  {\path{doi:10.1016/j.envsoft.2009.08.011}}.

\bibitem{Lupton18}
Richard~C. Lupton and Julian~M. Allwood.
\newblock {Incremental Material Flow Analysis with Bayesian Inference}.
\newblock {\em Journal of Industrial Ecology}, 22(6):1352--1364, 2018.
\newblock \href {https://doi.org/10.1111/jiec.12698}
  {\path{doi:10.1111/jiec.12698}}.

\bibitem{Dong23}
Jiayuan Dong, Jiankan Liao, Xun Huan, and Daniel Cooper.
\newblock Expert elicitation and data noise learning for material flow analysis
  using bayesian inference.
\newblock {\em Journal of Industrial Ecology}, 27(4):1105--1122, 2023.
\newblock URL:
  \url{https://onlinelibrary.wiley.com/doi/full/10.1111/jiec.13399}, \href
  {https://doi.org/https://doi.org/10.1111/jiec.13399}
  {\path{doi:https://doi.org/10.1111/jiec.13399}}.

\bibitem{Wang22}
Junyang Wang, Kolyan Ray, Pablo Brito-Parada, Yves Plancherel, Tom Bide, Joseph
  Mankelow, John Morley, Julia Stegemann, and Rupert Myers.
\newblock Bayesian material flow analysis for systems with multiple levels of
  disaggregation and high dimensional data, 2022.
\newblock URL: \url{https://arxiv.org/abs/2211.06178}.

\bibitem{Berger85}
James~O. Berger.
\newblock {\em {Statistical Decision Theory and Bayesian Analysis}}.
\newblock Springer Series in Statistics. Springer New York, New York, NY, 1985.
\newblock \href {https://doi.org/10.1007/978-1-4757-4286-2}
  {\path{doi:10.1007/978-1-4757-4286-2}}.

\bibitem{Jaynes03}
E.~T. Jaynes.
\newblock {\em Probability Theory: The Logic of Science}.
\newblock Cambridge University Press, 2003.
\newblock \href {https://doi.org/10.1017/CBO9780511790423}
  {\path{doi:10.1017/CBO9780511790423}}.

\bibitem{Sivia06}
D.~S. Sivia and J.~Skilling.
\newblock {\em {Data Analysis: A Bayesian Tutorial}}.
\newblock Oxford University Press, New York, NY, 2nd edition, 2006.

\bibitem{Bertsekas08}
D.P. Bertsekas and J.N. Tsitsiklis.
\newblock {\em Introduction to Probability}.
\newblock Athena {Scientific} Optimization and Computation Series. Athena
  Scientific, Nashua, NH, 2008.

\bibitem{VonToussaint11}
Udo {Von Toussaint}.
\newblock {Bayesian inference in physics}.
\newblock {\em Reviews of Modern Physics}, 83:943--999, 2011.
\newblock \href {https://doi.org/10.1103/RevModPhys.83.943}
  {\path{doi:10.1103/RevModPhys.83.943}}.

\bibitem{Anspach24}
R.~L. Anspach, S.~R. Allen, and R.~C. Lupton.
\newblock Robust modeling of material flows to end-uses under uncertainty: Uk
  wood flows and material efficiency opportunities.
\newblock {\em Journal of Industrial Ecology}, 28(4):953--965, 2024.
\newblock \href {https://doi.org/https://doi.org/10.1111/jiec.13511}
  {\path{doi:https://doi.org/10.1111/jiec.13511}}.

\bibitem{Bhuwalka23}
Karan Bhuwalka, Eunseo Choi, Elizabeth~A. Moore, Richard Roth, Randolph~E.
  Kirchain, and Elsa~A. Olivetti.
\newblock A hierarchical bayesian regression model that reduces uncertainty in
  material demand predictions.
\newblock {\em Journal of Industrial Ecology}, 27(1):43--55, 2023.
\newblock URL:
  \url{https://onlinelibrary.wiley.com/doi/full/10.1111/jiec.13339}, \href
  {https://doi.org/https://doi.org/10.1111/jiec.13339}
  {\path{doi:https://doi.org/10.1111/jiec.13339}}.

\bibitem{Klinglmair16}
Manfred Klinglmair, Ottavia Zoboli, David Laner, Helmut Rechberger,
  Thomas~Fruergaard Astrup, and Charlotte Scheutz.
\newblock The effect of data structure and model choices on mfa results: A
  comparison of phosphorus balances for denmark and austria.
\newblock {\em Resources, Conservation and Recycling}, 109:166--175, 2016.
\newblock URL:
  \url{https://www.sciencedirect.com/science/article/pii/S092134491630043X?casa_token=KDE8dtqrwokAAAAA:jdGt0JTaSmM4b41uH7buBGsudTtmA6GFHACjcN3H_NJmrYtAOXcoIWCzAgHilF2oM0lbKPg},
  \href {https://doi.org/https://doi.org/10.1016/j.resconrec.2016.03.009}
  {\path{doi:https://doi.org/10.1016/j.resconrec.2016.03.009}}.

\bibitem{Chatterjee23}
Abheek Chatterjee, Cade Helbig, Richard Malak, and Astrid Layton.
\newblock A comparison of graph-theoretic approaches for resilient system of
  systems design.
\newblock {\em Journal of Computing and Information Science in Engineering},
  23(3), 2023.
\newblock URL:
  \url{https://asmedigitalcollection.asme.org/computingengineering/article/23/3/030906/1160385/A-Comparison-of-Graph-Theoretic-Approaches-for?casa_token=s4AaGCIICLIAAAAA:4pxlgdWPZj-E7YhqpKoqKQER80X_8JZt23TIFg_vp8rAnfLQnJhn1kr7ET8Gyjpi6ivfTQ},
  \href {https://doi.org/https://doi.org/10.1115/1.4062231}
  {\path{doi:https://doi.org/10.1115/1.4062231}}.

\bibitem{Ulanowicz09}
Robert~E. Ulanowicz, Sally~J. Goerner, Bernard Lietaer, and Rocio Gomez.
\newblock Quantifying sustainability: Resilience, efficiency and the return of
  information theory.
\newblock {\em Ecological Complexity}, 6(1), 2009.
\newblock URL:
  \url{https://www.sciencedirect.com/science/article/pii/S1476945X08000561?casa_token=A6L_XpLomo0AAAAA:Z2eOc7TbPNkELuF9aXSv3n8Rfw8vIRwzBEJKbFr8ROSwysVuv9h1lE06eURdga1KvcY9xzk},
  \href {https://doi.org/https://doi.org/10.1016/j.ecocom.2008.10.005}
  {\path{doi:https://doi.org/10.1016/j.ecocom.2008.10.005}}.

\bibitem{Schwab16}
Oliver Schwab, David Lanar, and Helmut Rechberger.
\newblock Quantitative evaluation of data quality in regional material flow
  analysis.
\newblock {\em Journal of Industrial Ecology}, 6(1), 2016.
\newblock URL: \url{https://onlinelibrary.wiley.com/doi/10.1111/jiec.12490},
  \href {https://doi.org/https://doi.org/10.1111/jiec.12490}
  {\path{doi:https://doi.org/10.1111/jiec.12490}}.

\bibitem{Cullen12}
Jonathan~M. Cullen, Julian~M. Allwood, and Margarita~D. Bambach.
\newblock {Mapping the Global Flow of Steel: From Steelmaking to End-Use
  Goods}.
\newblock {\em Environmental Science \& Technology}, 46(24):13048--13055, 2012.
\newblock \href {https://doi.org/10.1021/es302433p}
  {\path{doi:10.1021/es302433p}}.

\bibitem{OHagan05}
Paul~H Garthwaite, Joseph~B Kadane, and Anthony O'Hagan.
\newblock Statistical methods for eliciting probability distributions.
\newblock {\em Journal of the American Statistical Association},
  100(470):680--701, 2005.
\newblock \href
  {http://arxiv.org/abs/https://doi.org/10.1198/016214505000000105}
  {\path{arXiv:https://doi.org/10.1198/016214505000000105}}, \href
  {https://doi.org/10.1198/016214505000000105}
  {\path{doi:10.1198/016214505000000105}}.

\bibitem{Liu13}
Gang Liu, Colton~E. Bangs, and Daniel~B. Muller.
\newblock Stock dynamics and emission pathways of the global aluminium cycle.
\newblock {\em Nature Climate Change}, 3:338--342, 2013.
\newblock URL: \url{https://www.nature.com/articles/nclimate1698}, \href
  {https://doi.org/https://doi.org/10.1038/nclimate1698}
  {\path{doi:https://doi.org/10.1038/nclimate1698}}.

\bibitem{Graedel05}
T.~E. Graedel, Dick van Beers, Marlen Bertram, Kensuke Fuse, Robert~B. Gordon,
  Alexander Gritsinin, Ermelinda~M. Harper, Amit Kapur, Robert~J. Klee, Reid
  Lifset, Laiq Memon, and Sabrina Spatari.
\newblock The multilevel cycle of anthropogenic zinc.
\newblock {\em Journal of Industrial Ecology}, 9(3):67--90, 2005.
\newblock \href {https://doi.org/https://doi.org/10.1162/1088198054821573}
  {\path{doi:https://doi.org/10.1162/1088198054821573}}.

\bibitem{Bornhoft21}
Nikolaus~A. Bornhöft, Bernd Nowack, and Lorenz~M. Hilty.
\newblock Representation, propagation, and interpretation of uncertain
  knowledge in dynamic probabilistic material flow models.
\newblock {\em Environmental Modeling \& Assessment}, 26:709--721, 2021.
\newblock \href {https://doi.org/https://doi.org/10.1007/s10666-021-09775-5}
  {\path{doi:https://doi.org/10.1007/s10666-021-09775-5}}.

\bibitem{Ghosh58}
Amitav Ghosh.
\newblock Input-output approach in an allocation system.
\newblock {\em Economica}, 25(97):58--64, 1958.
\newblock \href {https://doi.org/https://doi.org/10.2307/2550694}
  {\path{doi:https://doi.org/10.2307/2550694}}.

\bibitem{Leontief36}
Wassily~W. Leontief.
\newblock Quantitative input and output relations in the economic systems of
  the united states.
\newblock {\em The Review of Economics and Statistics}, 18(3):105--125, 1936.
\newblock \href {https://doi.org/https://doi.org/10.2307/1927837}
  {\path{doi:https://doi.org/10.2307/1927837}}.

\bibitem{Jeffreys98}
Harold Jeffreys.
\newblock {\em The theory of probability}.
\newblock OuP Oxford, 1998.

\bibitem{Akaike74}
Hirotugu Akaike.
\newblock A new look at the statistical model identification.
\newblock {\em IEEE transactions on automatic control}, 19(6):716--723, 1974.

\bibitem{Schwarz78}
Gideon Schwarz.
\newblock Estimating the dimension of a model.
\newblock {\em The Annals of Statistics}, 6(2):461--464, 1978.

\bibitem{Kass1995}
Robert~E. Kass and Adrian~E. Raftery.
\newblock {Bayes Factor}.
\newblock {\em Journal of American Statistical Association}, 90(430):773--795,
  1995.
\newblock \href {https://doi.org/10.2307/2291091} {\path{doi:10.2307/2291091}}.

\bibitem{Wasserman2000}
Larry Wasserman.
\newblock {Bayesian Model Selection and Model Averaging}.
\newblock {\em Journal of Mathematical Psychology}, 44:92--107, 2000.
\newblock \href {https://doi.org/10.1006/jmps.1999.1278}
  {\path{doi:10.1006/jmps.1999.1278}}.

\bibitem{Friel2012}
Nial Friel and Jason Wyse.
\newblock {Estimating the evidence - a review}.
\newblock {\em Statistica Neerlandica}, 66(3):288--308, 2012.
\newblock \href {https://doi.org/10.1111/j.1467-9574.2011.00515.x}
  {\path{doi:10.1111/j.1467-9574.2011.00515.x}}.

\bibitem{Jeffreys35}
Harold Jeffreys.
\newblock Some tests of significance, treated by the theory of probability.
\newblock {\em Mathematical Proceedings of the Cambridge Philosophical
  Society}, 31(2):203--222, 1935.
\newblock \href {https://doi.org/https://doi.org/10.1017/S030500410001330X}
  {\path{doi:https://doi.org/10.1017/S030500410001330X}}.

\bibitem{Chopin2020}
Nicolas Chopin and Omiros Papaspiliopoulos.
\newblock {\em {An Introduction to Sequential Monte Carlo Methods}}.
\newblock Springer Nature Switzerland, Cham, Switzerland, 2020.
\newblock \href {https://doi.org/10.1046/j.1467-9884.2003.t01-6-00383_8.x}
  {\path{doi:10.1046/j.1467-9884.2003.t01-6-00383_8.x}}.

\bibitem{Drovandi14}
Christopher~C. Drovandi, James~M. McGree, and Anthony~N. Pettitt.
\newblock A sequential monte carlo algorithm to incorporate model uncertainty
  in bayesian sequential design.
\newblock {\em Journal of Computational and Graphical Statistics}, 23(1):3--24,
  2014.
\newblock \href {https://doi.org/https://doi.org/10.1080/10618600.2012.730083}
  {\path{doi:https://doi.org/10.1080/10618600.2012.730083}}.

\bibitem{Leontief86}
Wassily Leontief.
\newblock {\em Input-Output Economics}.
\newblock Oxford University Press, 1986.

\bibitem{Kitzes13}
Justin Kitzes.
\newblock An introduction to environmentally-extended input-output analysis.
\newblock {\em Resources}, 2(4):489--503, 2013.

\bibitem{McPhail18}
C.~McPhail, H.~R. Maier, J.~H. Kwakkel, M.~Giuliani, A.~Castelletti, and
  S.~Westra.
\newblock Robustness metrics: How are they calculated, when should they be used
  and why do they give different results?
\newblock {\em Earth's Future}, 6(2):169--191, 2018.
\newblock \href {https://doi.org/https://doi.org/10.1002/2017EF000649}
  {\path{doi:https://doi.org/10.1002/2017EF000649}}.

\bibitem{USGS12a}
{United States Geological Survey}.
\newblock Iron and steel, 2012.
\newblock
  \url{https://www.usgs.gov/centers/national-minerals-information-center/iron-and-steel-statistics-andinformation}.

\bibitem{USGS12b}
{United States Geological Survey}.
\newblock Iron and steel scrap, 2012.
\newblock
  \url{https://www.usgs.gov/centers/national-minerals-information-center/iron-and-steel-scrapstatistics-and-information}.

\bibitem{USGS12c}
{United States Geological Survey}.
\newblock Iron ore, 2012.
\newblock
  \url{https://www.usgs.gov/centers/national-minerals-information-center/iron-ore-statistics-andinformation}.

\bibitem{WSA12}
{World Steel}.
\newblock Steel statistical yearbook, 2012.
\newblock
  \url{https://worldsteel.org/wp-content/uploads/Steel-Statistical-Yearbook-2012.pdf}.

\bibitem{Blei17}
David~M. Blei, Alp Kucukelbir, and Jon~D. McAuliffe.
\newblock Variational inference: A review for statisticians.
\newblock {\em Journal of the American Statistical Association},
  112(518):859--877, 2017.

\bibitem{Daehn17}
Katrin~E Daehn, Andr{\'e} Cabrera~Serrenho, and Julian~M Allwood.
\newblock How will copper contamination constrain future global steel
  recycling?
\newblock {\em Environmental science \& technology}, 51(11):6599--6606, 2017.

\bibitem{WEF23}
{World Economic Forum}.
\newblock Closing the loop on automotive steel: A policy agenda, 2023.
\newblock
  \url{https://www.google.com/url?sa=t&source=web&rct=j&opi=89978449&url=https://www3.weforum.org/docs/WEF_Closing_Loop_Automotive_Steel_2023.pdf&ved=2ahUKEwjP84Gat82JAxWh5ckDHZkrMVMQFnoECA8QAQ&usg=AOvVaw3bQrsAd4TKHDEQpcar-fom}.

\bibitem{Chaloner95}
Kathryn Chaloner and Isabella Verdinelli.
\newblock Bayesian experimental design: A review.
\newblock {\em Statistical science}, pages 273--304, 1995.

\bibitem{Muller05}
Peter M{\"{u}}ller.
\newblock {Simulation Based Optimal Design}.
\newblock {\em Handbook of Statistics}, 25:509--518, 2005.

\bibitem{Ryan2016}
Elizabeth~G. Ryan, Christopher~C. Drovandi, James~M. Mcgree, and Anthony~N.
  Pettitt.
\newblock {A review of modern computational algorithms for Bayesian optimal
  design}.
\newblock {\em International Statistical Review}, 84(1):128--154, 2016.
\newblock \href {https://doi.org/10.1111/insr.12107}
  {\path{doi:10.1111/insr.12107}}.

\bibitem{Rainforth2023}
Tom Rainforth, Adam Foster, Desi~R Ivanova, and Freddie~Bickford Smith.
\newblock {Modern Bayesian experimental design}.
\newblock {\em Statistical Science}, 39(1):100--114, 2024.
\newblock \href {https://doi.org/10.1214/23-STS915}
  {\path{doi:10.1214/23-STS915}}.

\bibitem{Huan2024}
Xun Huan, Jayanth Jagalur, and Youssef Marzouk.
\newblock {Optimal experimental design: Formulations and computations}.
\newblock {\em Acta Numerica}, 33:715--840, jul 2024.
\newblock \href {https://doi.org/10.1017/S0962492924000023}
  {\path{doi:10.1017/S0962492924000023}}.

\bibitem{Liao25}
Jiankan Liao, Xun Huan, and Daniel Cooper.
\newblock Bayesian optimal experimental design for intelligent data collection
  in material flow analysis.
\newblock {\em submitted to CIRP LCE}, 2025.

\bibitem{USGS14}
USGS.
\newblock {\em 2014 Minerals Yearbook: Iron and Steel}.
\newblock U.S. Geological Survey (2014), 2014.

\bibitem{USGS16}
USGS.
\newblock Iron and steel, 2016.

\bibitem{AISI15}
USGS.
\newblock Aisi. profile, 2015.
\newblock Arlington. Virginia 2015, 10.

\bibitem{Omar11}
Mohammed~A Omar.
\newblock {\em The automotive body manufacturing systems and processes}.
\newblock John Wiley \& Sons, 2011.

\bibitem{WSA09}
{World Steel}.
\newblock Yield improvement in the steel industry: Working group report 2003 -
  2006, 2009.

\bibitem{Milford11}
Rachel~L Milford, Julian~M Allwood, and Jonathan~M Cullen.
\newblock Assessing the potential of yield improvements, through process scrap
  reduction, for energy and co2 abatement in the steel and aluminium sectors.
\newblock {\em Resources, Conservation and Recycling}, 55(12):1185--1195, 2011.

\bibitem{Heidari24}
S.~M. Heidari, Y.~Zhu, A.~Tsai, G.~Keoleian, H.~C. Kim, R.~De~Kleine, J.~Kelly,
  S.~Shukla, and D.~Cooper.
\newblock Circular economy pathways for decarbonizing automotive aluminum and
  steel sheet components in the united states.
\newblock {\em Submitted to Journal of Sustainable Materials and Technologies},
  2024.

\bibitem{Zhu23}
Yongxian Zhu, Gregory~A Keoleian, and Daniel~R Cooper.
\newblock A parametric life cycle assessment model for ductile cast iron
  components.
\newblock {\em Resources, Conservation and Recycling}, 189:106729, 2023.

\bibitem{ANL23}
{Argonne National Laboratory}.
\newblock Greet model, 2023.
\newblock Retrieved from \url{https://www.anl.gov/topic/greet}.

\end{thebibliography}
\bibliographystyle{unsrturl}

\end{document}